\documentclass[twocolumn]{aastex62}
\usepackage{amsmath}
\usepackage{datetime}
\usepackage{appendix}
\usepackage{comment}
\usepackage{chngcntr}
\usepackage{soul}
\usepackage{xcolor}
\def\msun{{\rm M}_\odot}
\def\mpc{\rm Mpc}

\def\H2{\rm H_2}
\newcommand{\HI}{\rm H\,{\textsc i}} % HI with correct lettering

\newdateformat{specialdate}{\THEYEAR-\twodigit{\THEMONTH}-\twodigit{\THEDAY}}

\begin{document}

\defcitealias{Zhang2019}{Z19}
\defcitealias{Zhang2021}{Z21}

\title{Cold Gas in Massive Galaxies as A Critical Test of Black Hole Feedback Models}

\author[0000-0001-9879-4926]{Jingjing~Shi}
\affiliation{Kavli Institute for Astronomy and Astrophysics, Peking University, Beijing 100871, China}
\affiliation{Kavli Institute for the Physics and Mathematics of the Universe (WPI), The University of Tokyo Institutes for Advanced Study (UTIAS),
The University of Tokyo, 5-1-5 Kashiwanoha, Kashiwa-shi, Chiba, 277-8583, Japan}

\author{Yingjie~Peng}
\affiliation{Kavli Institute for Astronomy and Astrophysics, Peking University, Beijing 100871, China}

\author{Benedikt~Diemer}
\affiliation{Department of Astronomy, University of Maryland, College Park, MD 20742, USA}

\author{Adam~R.~H.~Stevens}
\affiliation{International Centre for Radio Astronomy Research, The University of Western Australia, Crawley, WA 6009, Australia}
\affiliation{Australian Research Council Centre of Excellence for All Sky Astrophysics in 3 Dimensions (ASTRO 3D)}

\author{Annalisa~Pillepich}
\affiliation{Max-Planck-Institut f\"{u}r Astronomie, K\"{o}nigstuhl 17, 69117 Heidelberg, Germany}

\author{Alvio~Renzini}
\affiliation{INAF - Osservatorio Astronomico di Padova, Vicolo dell’Osservatorio 5, I-35122 Padova, Italy}

\author{Jing~Dou}
\affiliation{Shanghai Astronomical Observatory, Chinese Academy of Sciences, 80 Nandan Road, Shanghai 200030, China}
\affiliation{University of Chinese Academy of Sciences, 19A Yuquan Road, Beijing 100049, People’s Republic of China}

\author{Yu~Gao}
\affiliation{Department of Astronomy, Xiamen University, Xiamen, Fujian 361005, China}
\affiliation{Purple Mountain Observatory \& Key Laboratory for Radio Astronomy, Chinese Academy of Sciences, 10 Yuanhua Road, Nanjing
210033, PR China}

\author{Qiusheng~Gu}
\affiliation{School of Astronomy and Space Science, Nanjing University, Nanjing 210093, China}
\affiliation{Key Laboratory of Modern Astronomy and Astrophysics (Nanjing University), Ministry of Education, Nanjing 210093, China}

\author{Luis~C.~Ho}
\affiliation{Kavli Institute for Astronomy and Astrophysics, Peking University, Beijing 100871, China}
\affiliation{Department of Astronomy, School of Physics, Peking University, 5 Yiheyuan Road, Beijing 100871, P. R. China}

\author{Xu~Kong}
\affiliation{CAS Key Laboratory for Research in Galaxies and Cosmology, Department of Astronomy, University of Science and Technology of China, Hefei 230026, China}
\affiliation{School of Astronomy and Space Sciences, University of Science and Technology of China, Hefei 230026, China}

\author{Claudia~del~P.~Lagos}
\affiliation{International Centre for Radio Astronomy Research, The University of Western Australia, Crawley, WA 6009, Australia}
\affiliation{Australian Research Council Centre of Excellence for All Sky Astrophysics in 3 Dimensions (ASTRO 3D)}

\author{Di~Li}
\affiliation{CAS Key Laboratory of FAST, National Astronomical Observatories, Chinese Academy of Sciences, Beijing 100012, P. R. China}
\affiliation{School of Astronomy and Space Science, University of Chinese Academy of Sciences, Beijing 100049, P. R. China}

\author[0000-0001-9592-4190]{Jiaxuan~Li}
\affiliation{Kavli Institute for Astronomy and Astrophysics, Peking University, Beijing 100871, China}
\affiliation{Department of Astrophysical Sciences, 4 Ivy Lane, Princeton University, Princeton, NJ 08544, USA}

\author{Roberto~Maiolino}
\affiliation{Cavendish Laboratory, University of Cambridge, 19 J. J. Thomson Avenue, Cambridge CB3 0HE, UK}
\affiliation{Kavli Institute for Cosmology, University of Cambridge, Madingley Road, Cambridge CB3 0HA, UK}
\affiliation{Department of Physics and Astronomy, University College London, Gower Street, London WC1E 6BT, UK}

\author{Filippo~Mannucci}
\affiliation{Istituto Nazionale di Astrofisica, Osservatorio Astrofisico di Arcetri, Largo Enrico Fermi 5, I-50125 Firenze, Italy}

\author{Lizhi~Xie}
\affiliation{Tianjin Normal University, Binshuixidao 393, Tianjin, China}

\author{Chengpeng~Zhang}
\affiliation{Kavli Institute for Astronomy and Astrophysics, Peking University, Beijing 100871, China}

\correspondingauthor{Jingjing~Shi \& Yingjie~Peng}
\email{jingssrs1989@gmail.com; yjpeng@pku.edu.cn}

%%%%%%%%%%%
%Abstract
%%%%%%%%%%%
\begin{abstract}
Black hole feedback has been widely implemented as the key recipe to quench star formation in massive galaxies in  modern semi-analytic models and hydrodynamical simulations. 
As the theoretical details surrounding the accretion and feedback of black holes continue to be refined, various feedback models have been implemented across simulations, with notable differences in their outcomes.
Yet, most of these simulations have successfully reproduced some observations, such as stellar mass function and star formation rate density in the local Universe.  We use the recent observation on the change of neutral hydrogen gas mass (including both $\H2$ and $\HI$) with star formation rate of massive central disc galaxies
as a critical constraint of black hole feedback models across several simulations. We find that the predictions of IllustrisTNG agree with the observations much better than the other models tested in this work. This favors IllustrisTNG's treatment of active galactic nuclei - where kinetic winds are driven by black holes at low accretion rates - as more plausible amongst those we test. In turn, this also indirectly supports the idea that the massive central disc galaxy population in the local Universe was likely quenched by AGN feedback.
\end{abstract}

\keywords{methods: numerical - galaxies: evolution  - galaxies: star formation - galaxies: ISM}
%%%%%%%%%%%%%%%%%%%%%%%%%%%%%%
%introduction
%%%%%%%%%%%%%%%%%%%%%%%%%%%%%%
\section{Introduction}
\label{sec_intro}

Baryons cool and form stars within the potential well of dark matter halos \citep{ReesOstriker1977,WhiteRees1978,WhiteFrenk1991}. However, only $5\%$--$25\%$ of baryons within dark matter halos are converted into stars by $z=0$ efficiently in most galaxies \citep{WechslerTinker2018,Dutton2010,Zheng2007,Moster2018,WangJing2010,Yang2009,Yang2012,Kravtsov2018,Behroozi2019}.
Stellar feedback, such as supernova (SN) explosions and stellar winds, can heat up and even eject the gas in low-mass systems, reducing their star formation (SF) efficiency \citep{Larson1974,WhiteRees1978,Silk2003,Springel2003}. However, to suppress the SF efficiency in massive systems where the potential well is deeper, SN feedback alone is not enough \citep{Benson2003}. Active galactic nuclear (AGN) feedback is invoked as a necessary mechanism to reduce further SF activity in massive systems in galaxy formation and evolution models (\citealt{Silk1998,Croton2006,Bower2006,Henriques2015,2014ARA&A..52..529Y,2015ApJ...804..101Y,2018ApJ...857..121Y}, see \citealt{SomervilleDave2015} and \citealt{NaabOstriker2017} for a review on the current status of galaxy formation and evolution models).   

AGN feedback is launched in quite different ways in different theoretical models of cosmological simulation and semi-analytic models (SAMs). In general, the way that feedback is launched depends on the BH accretion rate: the high accretion rate mode and low accretion rate mode. In SAMs \citep{Croton2006,Bower2006,Guo2011,Henriques2015}, BHs continue to accrete gas from the circumgalactic medium, the induced feedback injects thermal energy into the gas in DM halo to effectively reduce the hot-gas cooling rate. In Illustris \citep{Vogelsberger2013,Torrey2014} and IllustrisTNG \citep{Weinberger2017,Pillepich2018a}, when the BH accretion rate is higher than a certain threshold, the thermal energy is inserted into the surrounding gas within galaxies; when the BH accretion rate is lower, a hot bubble is injected (in Illustris) or a certain amount of kinetic energy is added to the gas surrounding the BH (so called "kinetic feedback", in IllustrisTNG). In EAGLE, a single heating mode from BH is launched, which nevertheless mimics the relatively quiescent ‘radio mode’ and vigorous ‘quasar mode’ when the BH accretion rate is much smaller than or similar to the Eddington rate \citep{Schaye2015,Crain2015}. Despite the non-negligible differences among these implementations, they are all adjusted to reproduce the $z=0$ stellar mass function and the stellar mass--BH mass scaling relation to various levels of agreement. More observations, especially those that are not implemented for model calibration, are needed to constrain the AGN feedback models in cosmological simulations and SAMs.

One promising constraint comes from investigating the gas content in galaxies. AGN feedback can act either negatively or positively \citep{Cresci2018}. Negative AGN feedback is believed to suppress the SF in massive galaxies in two ways: through prevention \citep{Cresci2015} and ejection. Preventative feedback is when the hot gas can not cool efficiently due to the heating from AGN, while ejective feedback is when cold gas is pushed away from the galaxy due to the wind produced by the accreting BH \citep{Zinger2020}. The role of AGN feedback being preventive or ejective is not necessarily directly tied to how the feedback is implemented in practice. For example, the kinetic mode in IllustrisTNG can be also preventive, as kinetic energy can also transform into thermal energy via shocks \citep{Weinberger2017}; likewise, the thermal injection in EAGLE can produce gas ouflows (i.e. can be ejective too) due to the induced pressure gradients.
By contrast, and of lesser importance, positive AGN feedback is when additional star formation is induced by the compression of molecular gas in a galaxy's disc \citep{Silk2013} or directly in the outflowing cold gas \citep{Ishibashi2012,Zubovas2013,Maiolino2017,Gallagher2019}. Understanding how AGN feedback impacts galaxies' gas and star formation properties can help us to distinguish/constrain various implementations in these models \citep{Terrazas2020}.

On average, passive galaxies unsurprisingly have less cold gas than star-forming ones \citep{Saintonge2016,Tacconi2018,Catinella2018}.
However, when massive central disc galaxies ($10^{10.6}<M_\star/{\rm M}_{\odot}<10^{11}$) are selected, i.e.~focusing on internal quenching mechanisms and excluding external environmental ones, \cite{Zhang2019} (\citetalias{Zhang2019} hereafter) found that, as SFR decreases, their $\HI$ gas mass remains surprisingly constant, but both $\H2$ gas mass and $\H2$ star formation efficiency decrease. \cite{Zhang2021} ((\citetalias{Zhang2021} hereafter)) further show the change of gas content is also accompanied by the rapid increase of stellar concentration index,
bulge-to-total mass ratio, central velocity dispersion and AGN frequency (which in \citetalias{Zhang2019} are mostly LINERs). These altogether suggest more massive black holes, and possibly stronger AGN feedback, in quenching or quenched galaxies. These results are robust against different SFR estimators, including the ${\rm H}{\alpha}$/$D_{4000}$-based SFR \citep{Brinchmann2004} with aperture correction by performing spectral energy distribution (SED) fitting to the photometry outside the fiber, and SFRs obtained from SED fitting of UV, optical, and mid-IR bands \citep{Salim2016,Salim2018}, see more detailed discussion in \citetalias{Zhang2021}. Since cold gas is the fuel of star formation and can provide direct evidence of how quenching may happen in galaxies, this observational result of massive central disc galaxies can be used as a new constraint for current galaxy formation and evolution simulations/models, especially for AGN feedback. In particular, since $\H2$ is mainly distributed in the inner stellar disc and $\HI$ is usually located at larger radii, the observed $\HI$ and $\H2$ versus SFR relations as in \citetalias{Zhang2019} potentially put a strong observational constraint on the strength of AGN feedback and on its ``inside-out" nature, as it needs to clear out most of the $\H2$ gas in the inner disc while retaining much of the $\HI$ gas in the outer disc.

In this work, we focus mainly on the prediction for the neutral hydrogen gas--SFR relation of massive central discs produced by IllustrisTNG, EAGLE, L-Galaxies, and Illustris. We show that IllustrisTNG agrees with the observed gas--SFR relation better than the others.
We then study the gas distribution and quenching mechanism of the central discs in IllustrisTNG in more detail, as it may give useful clues on the quenching process in the real Universe.

%%%%%%%%%%%
%Data and Methods
%%%%%%%%%%%
\section{Data and methods}
\label{sec_method}
\subsection{The IllustrisTNG simulation}
The IllustrisTNG project \citep{Springel2018,Marinacci2018,Naiman2018,Pillepich2018b,Nelson2018} is a suite of cosmological hydrodynamical simulations in a $\Lambda$CDM Universe run with the moving-mesh code {\sc arepo} \citep{Springel2010}. It contains simulations of three different volumes, TNG50, TNG100, and TNG300 with respective box lengths of $35\,h^{-1}\,\mpc$, $75\,h^{-1}\,\mpc$, and $205\,h^{-1}\,\mpc$. In this work, we use the publicly available TNG100 data as a good balance between resolution and volume \citep{Nelson2019}\footnote{\url{https://www.tng-project.org}}. The numerical resolution of TNG100 is also closest to the test runs on which the free parameters of the subgrid physics were calibrated, which results in the best match with observational constraints (e.g. stellar mass function etc.).
The IllustrisTNG galaxy formation and evolution model includes gas cooling and heating, star formation, stellar evolution and chemical enrichment, SN feedback, BH growth, AGN feedback, and cosmic magnetic field (see \citealt{Weinberger2017, Pillepich2018a} for more detailed information on the models).

Galaxies in TNG100 are identified using the Friends-of-Friends (FOF) and {\tt SUBFIND} algorithm \citep{Davis1985,Springel2001}. The neutral gas fraction of non-star-forming gas cells is calculated self-consistently in the simulation by computing the cooling rate and photo-ionization rate due to the UV background, while star-forming gas cells are modelled as a two-phase medium following \cite{Springel2003}, where the hot phase gas is assumed fully ionized and the cold phase gas fully neutral (see appendix A1 of \citealt{Stevens2019a} for more information).
In each gas cell, the molecular hydrogen fraction can be obtained by post-processing using empirical, simulation-based, or theoretical prescriptions. In this work, we use the molecular hydrogen gas fraction calculated using five different models \citep{Diemer2018}: the empirical model where the molecular gas fraction is found to be correlated with the mid-plane pressure (\citealt{Leroy2008}, L08 hereafter), the high-resolution chemical-network-inclusive simulations that produced calibrated relations between molecular gas fraction and surface density, metallicity, and UV field (\citealt{Gnedin2011}, GK11 hereafter; \citealt{Gnedin2014}, GD14 hereafter), and analytical equilibrium models of molecular clouds with detailed calculations of molecular hydrogen creation and destruction (\citealt{Krumholz2013}, K13 hereafter; \citealt{Sternberg2014}, S14 hereafter). Galaxies are rotated into a face-on position using the angular momentum of gas and stars, and the $\HI$ and ${\rm H_2}$ fractions are computed using the projected quantities (such as surface density, SFR surface density etc.) That is, we use the `map' methods of \citet{Diemer2018}. We refer the readers to \cite{Diemer2018,Diemer2019,Stevens2019a,Stevens2019b,Stevens2021} for more details on the decomposition methods and comparison with observation (also see \citealt{Dave2020} for an additional comparison with other hydrodynamic simulations). 

In this work, $M_{\star}$ for each galaxy is calculated within the radius of $2R_{\star}$, where $R_{\star}$ is the stellar half-mass radius. To approximate the scale of the Arecibo beam width ($3.4'$ at $21$\,cm, e.g. \citealt{Jones2018}) for the relevant galaxies in the ALFALFA survey, we only count the $\HI$ gas within a fixed radius of $70$ kpc (i.e. the corresponding physical scale of the beam witdth at $z=0.037$) as in \cite{Diemer2019} and \cite{Bahe2016}; but see \citet{Stevens2019a} for a more precise method. We also test our results with fixed radii of $50$ kpc and $80$ kpc and see no significant variations of our results. We also count the $\H2$ gas within the radius of $70$ kpc and test the results using the radius of $20$ kpc (which roughly corresponds to the IRAM aperture for xCOLD GASS survey). Our results stay unchanged since $\H2$ is mainly located in the central region of galaxies.
To make a direct comparison with the observed trend of cold gas content versus SFR as in \citetalias{Zhang2019} and \citetalias{Zhang2021}, we apply the same selection criteria of massive central disc galaxies with $10^{10.6}<M_\star/{\rm M}_{\odot}<10^{11}$. We use the $\kappa_{\rm rot}$ parameter to distinguish disc galaxies from elliptical galaxies, defined as the ratio of the ordered rotation versus the total kinetic energy using stellar particles \citep{Sales2012,Rodriguez-Gomez2015}. If $\kappa_{\rm rot}>0.5$, the galaxy is rotation-dominated, disc-like; if $\kappa_{\rm rot}<0.5$, the galaxy is dispersion-dominated, elliptical-like. Figure A1 in \cite{Diemer2019} compared the elliptical galaxy fraction in TNG100 based on $\kappa_{\rm rot}$ with the observed fraction given by \cite{Calette2018}, finding a good match between simulations and observations in the stellar mass range studied in this work. In Section~\ref{sec_res} and Appendix~\ref{sec_morph} we give more discussion on the morphology selection. With the above selection, our final sample in TNG100 includes $674$ centrals, of which $340$ are disc-like. 
The SFRs are obtained by averaging the star formation over the past $200\, {\rm Myr}$ within two different apertures: everything that are gravitation-ally self-bounded within the SUBFIND identified subhalo, SFR(all grav.), and that within $2R_\star$, SFR($<2R_\star$). We take SFR(all grav.) as our default choice. In Appendix \ref{sec_SFR}, we also test our results with SFR calculated using various apertures and time-scales, and find that the general trends remain similar. 

Due to the finite mass resolution in the simulation, galaxies' SFRs suffer a resolution limit (see \citealt{Donnari2019} for a detailed discussion). For SFRs averaged in the past $200\, {\rm Myr}$, this SFR resolution limit is $10^{-2.46}\,{\rm M}_{\odot}\,{\rm yr^{-1}}$. For galaxies with SFRs below this limit, we reassign each of them a random SFR between $10^{-2.8}$--$10^{-2.5}\,{\rm M}_{\odot}\,{\rm yr^{-1}}$ so that they show up in the log-scale scatter plots.

\subsection{L-Galaxies, EAGLE, and Illustris-1}
L-Galaxies 2020 is the newest version of the Munich semi-analytic galaxy formation model \citep{Henriques2020}\footnote{\url{https://lgalaxiespublicrelease.github.io/}}. This model is built upon the subhalo merger trees from the Millennium and Millennium-II simulations scaled to first-year {\it Planck} cosmology. The evolution of baryonic components are traced in the model, including hot gas atmosphere, cold interstellar gas, a gas reservoir ejected by winds, stars in the bulge, disc, and intracluster light, and central supermassive BHs. Elemental abundances (including H) are traced in a galactic chemical enrichment scheme introduced by \citet{Yates2013}. In this work, we select the central galaxies with $10^{10.6}<M_\star/\msun<10^{11}$ and disc-to-total stellar mass ratios (D/T) $>0.5$ from the publicly available online database\footnote{\url{http://www.mpa-garching.mpg.de/millennium}}. The amount of hydrogen in cold gas component gives us the neutral hydrogen mass.

The EAGLE simulation \citep{Schaye2015,Crain2015,McAlpine2016}\footnote{\url{http://eagle.strw.leidenuniv.nl}} is a cosmological hydrodynamical simulation run with a smoothed particle hydrodynamics' (SPH) solver \textsc{gadget3} with {\it Planck} cosmology. We use the largest-volume run, Ref-L100N1504, with box length of $100$ Mpc. Halos and galaxies are identified using the FOF and {\tt SUBFIND} algorithms. Stochastic, isotropic heating of gas particles is implemented for SF feedback ($\Delta T_{\rm SF}=10^{7.5}$K) and BH feedback ($\Delta T_{\rm BH}=10^{8.5}$K), where $\Delta T_{\rm SF}$ and $\Delta T_{\rm BH}$ are the temperature jump of gas particles receiving feedback energy. The neutral hydrogen fractions of gas particles are estimated in post-processing \citep{Lagos2015,Bahe2016,Crain2017}, using the fitting function of \cite{Rahmati2013}, which is calibrated using TRAPHIC radiative transfer simulations \citep{Pawlik2008}. \cite{Dave2020} compared the atomic and molecular hydrogen content of galaxies from IllustrisTNG, EAGLE, and SIMBA \citep{Dave2019}, we refer the readers to there for more information. Similar to TNG100, the massive central discs are selected using $\kappa_{\rm rot}>0.5$. The neutral hydrogen mass are calculated within a radius of $60$ kpc, however, the gas content stays roughly unchanged even with a different radius choice of $40$ kpc.

The Illustris simulation is the precursor simulation of IllustrisTNG \citep{Vogelsberger2014,Sijacki2015,Genel2014,Nelson2015}\footnote{\url{https://www.illustris-project.org}}. In this work, we use Illustris-1, which has the same initial condition, volume, and resolution as TNG100. The main difference of Illustris, compared to IllustrisTNG, is the treatment of BH accretion and feedback (as discussed in Section~\ref{sec_intro}) and of galactic winds. Illustris also lacks magnetohydrodynamics. The central discs in Illustris-1 are selected in the same way as TNG100. The neutral hydrogen gas mass of a galaxy is also summed up over the gas cells within the radius of $70$ kpc.

An non-trivial caveat in the above models is that none of them attempts to actually resolve the neutral phases of the ISM. Thus the predicted neutral hydrogen fraction is not a real prediction based on the ISM chemistry, but a modeling related to the effective equation of state function, where star formation is initiated once the gas density $n>0.13$cm$^{-3}$.

%%%%%%%%%%%
%Results
%%%%%%%%%%%
\section{Results} 
\label{sec_res}

\subsection{Total neutral hydrogen in simulations and real galaxies}

\begin{figure}
\centering 
 \includegraphics[width=1.\linewidth]{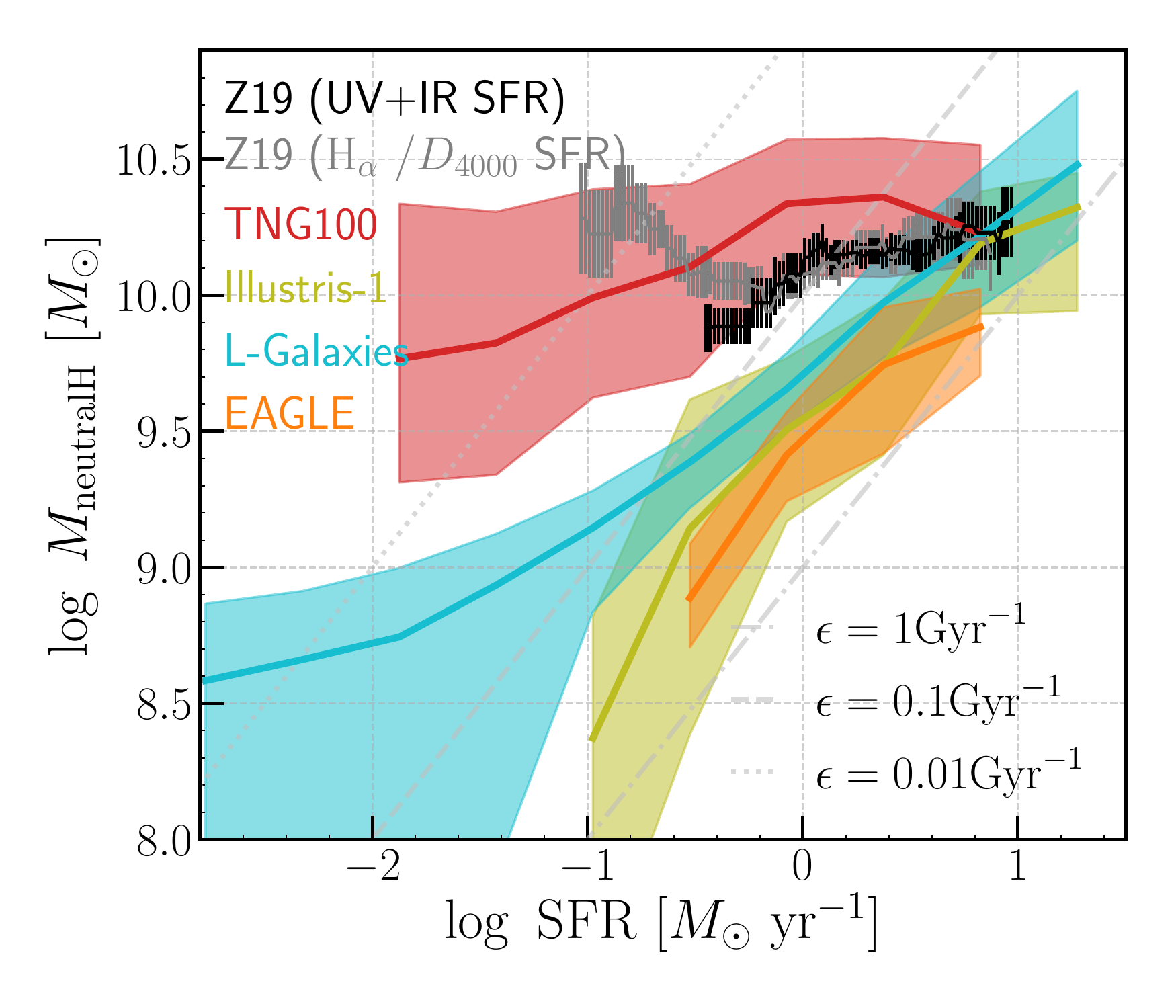}
\caption{\label{fig_mnh_sfr} Neutral hydrogen ($\HI+\H2$) mass versus SFR for central disc galaxies ($10^{10.6}<M_\star/{\rm M}_{\odot}<10^{11}$) in TNG100 ($\kappa_{\rm rot}>0.5$), Illustris ($\kappa_{\rm rot}>0.5$), L-Galaxies (D/T$>0.5$), and EAGLE ($\kappa_{\rm rot}>0.5$). The black and gray lines show the observational results from \citetalias{Zhang2019} for SFRs calculated using ${\rm H\alpha}$ emission and $D_{4000}$ (corrected for the limited SDSS fiber aperture) and UV+IR SEDs, respectively. The solid lines are the median relations, with the 16th--84th percentile ranges shown with the shaded regions. We also overplot lines of fixed star formation efficiency $\epsilon={\rm SFR}/M_{\rm gas}$ for $1\,{\rm Gyr^{-1}}$, $0.1\,{\rm Gyr^{-1}}$, and $0.01\,{\rm Gyr^{-1}}$.}
\end{figure}

Figure \ref{fig_mnh_sfr} shows the neutral hydrogen (${\HI + \H2}$) gas mass versus SFR for massive central disc galaxies ($10^{10.6}<M_\star/{\rm M}_\odot<10^{11}$) in TNG100, EAGLE, Illustris, L-Galaxies, and observations \citepalias{Zhang2019}. In those four models, notably different black hole feedback models were implemented.

First, all simulations except for EAGLE reproduce well the observed neutral hydrogen gas content for the galaxies with largest SFRs, while EAGLE predicts about $0.3$ dex less gas than observed.
With SFR decreasing, only TNG100 roughly reproduces the \citetalias{Zhang2019} observational trend (which is also shown by the results in Figure \ref{fig_mhi_sfr_disc}), although the scatter around the median relation is not very small. By contrast, galaxies in EAGLE, L-Galaxies, and Illustris are too gas-poor at fixed SFR. When $10^{-0.5}<{\rm SFR} / ({\rm M}_\odot {\rm yr^{-1}})<10^{0.5}$, the neutral hydrogen gas decreases too fast in EAGLE, L-Galaxies, and Illustris, compared to observations and TNG100. For ${\rm SFR<10^{-0.5}\,{\rm M}_\odot\, {\rm yr^{-1}}}$, the neutral hydrogen gas keeps decreasing rapidly in EAGLE, Illustris, and L-Galaxies, and becomes significantly below the observed values (with either the H$\alpha$/$D_{4000}$ or UV+IR SFR estimator). 
We also explored an additional semi-analytic model, Shark \citep{2018MNRAS.481.3573L}, and found it behaves similarly to Illustris and EAGLE (not shown here to avoid overcrowding of lines).
Since the change in gas as SFR decreases in simulations is mainly driven by the implemented AGN feedback model for galaxies in this stellar mass range, the results in Figure \ref{fig_mnh_sfr} suggest that the AGN feedback during quenching might be too efficient in heating/expelling gas from the galaxies in EAGLE, L-Galaxies, and Illustris.

There is a concern on the \citetalias{Zhang2019} observation result that if different SFR estimator (UV+IR instead of ${\rm H}_{\alpha}/D_{\rm 4000}$) is used, 
the lowest SFRs of central disks are around log SFR=$-0.5$ whereas they go down to $\sim$ $-1$ in the former case. So, these galaxies appear having more or less progressed towards full quenching, depending on the adopted SFR diagnostics (see Figure 1 in \citealt{Cortese2020}, hereafter C20).
The effect of using alternative SFR estimator has been discussed in detail in \citetalias{Zhang2021}. The general trends of gas content and other galaxy properties with respect to SFR hold for both SFR estimators. It should also be noted that ``disc" galaxies are defined in a very different way in \citetalias{Zhang2019}\&\citetalias{Zhang2021} (visual classification by Galaxy Zoo) and C20 (a threshold in B/T). These two selections result in very different populations. As shown in \citetalias{Zhang2021}, the visually defined disc galaxies with low SFR often have massive bulges, and many of them will be classified as ellipticals using B/T; while most S0s have been classified as ellipticals (or uncertains) in Galaxy Zoo, but will be classified as discs by B/T. This different disc selection leads to the fact that the UV+IR derived SFRs in Z19 do not extend to the SFRs as low as in \citet{Cortese2020}. 

One issue in observation is that the SFRs derived from the SED fitting of UV, optical and mid-IR bands \citep{Salim2016,Salim2018} used in \citetalias{Zhang2019} and \citetalias{Zhang2021}, and the UV+IR SFRs used in \citet{Cortese2020} are not corrected for AGN contamination or by hot old stars. \citetalias{Zhang2021} shows almost 100\% of the massive discs with the lowest SFRs (by either SFR estimator) host LINERs. 
Thus the UV+IR SFRs are likely to be the upper limits and true SFRs could reach lower SFRs. One could subtract off an AGN component by doing the SED decomposition, but this is a very uncertain procedure. For these LINERs, \citetalias{Zhang2019} and \citetalias{Zhang2021} used the SFRs derived from $D_{\rm 4000}$, while such derivation is calibrated with non-AGN galaxies \citep{Brinchmann2004}. Apparently, SFRs from both estimators contain uncertainties at the low SFR end (dominated by LINERs). Therefore, in our comparison analysis, we use the results derived from both SFR estimators. As shown in Figure~\ref{fig_mnh_sfr}, the approximate agreement between TNG100 and either SFR estimator holds.

\subsection{$\HI$ and $\H2$ contents (in TNG100)}
\subsubsection{$M_{\HI}$ and $M_{\H2}$ versus SFR}

\begin{figure*}
\centering
 \includegraphics[width=1.\linewidth]{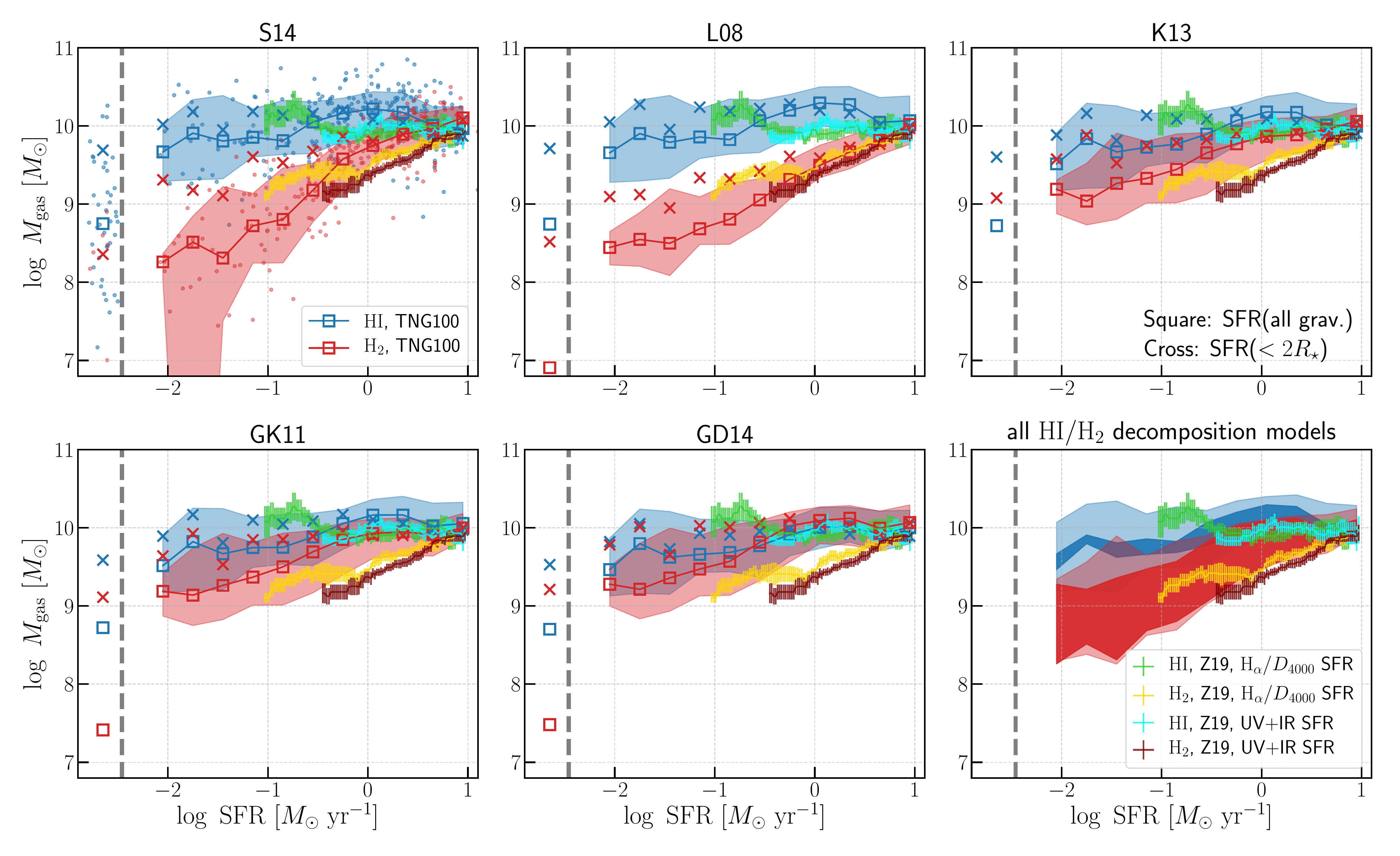}
\caption{\label{fig_mhi_sfr_disc} Atomic ($\HI$, blue) and molecular ($\H2$, red) hydrogen mass versus SFR in massive central disc galaxies in TNG100. The squares show the median relation for a 200-Myr historical SFR that is summed up over all the star particles gravitationally bound to a galaxy, i.e. to the subhalo; the crosses are the median relation based on particles within $2R_\star$.
The shaded areas indicate the 16th--84th percentile ranges. The first five panels show the results from five $\HI$/${\rm H_{2}}$ decomposition models as discussed in Section \ref{sec_method}. The small dots in the first panel show the distribution of individual galaxies in this massive central disc sample, and are not shown in other panels for clarity.
The lime-green ($\HI$) and gold ($\H2$) lines are the sliding medians from the observational results of \citetalias{Zhang2019}, where SFR is calculated using aperture corrected ${\rm H\alpha}$/$D_{4000}$ emission from the SDSS fibre spectra. The observational results using SFRs calculated from SED fitting of UV+IR bands (i.e.~Figure 6 of \citetalias{Zhang2019}) are overplotted.
The gray vertical dashed lines are the SFR resolution limits. 
We also overplot lines of fixed star formation efficiency $\epsilon={\rm SFR}/M_{\rm gas}$ for $1\,{\rm Gyr^{-1}}$, $0.1\,{\rm Gyr^{-1}}$, and $0.01\,{\rm Gyr^{-1}}$.   
In the last panel, the light blue and red shaded areas show the 16th--84th percentile range of the $\HI$ and $\H2$ gas components from all decomposition models for SFR(all grav.), and the darker areas indicate the range of the median relations of the five models.}
\end{figure*}

Given that TNG100 agrees the best with the observed neutral hydrogen mass--SFR relation, in Figure \ref{fig_mhi_sfr_disc} we show the $\HI$ and $\H2$ gas mass as a function of SFR for these massive central disc galaxies in TNG100, for the five different $\HI$/${\rm H_{2}}$ decomposition models as described in Section~\ref{sec_method}. In each panel, we show the results for SFRs derived within the entire subhalo (square) and $2R_\star$ (cross) to check the possible influence of varying apertures.
It is evident from Figure \ref{fig_mhi_sfr_disc} that above the SFR resolution limits (the vertical dashed line in each panel), all five of these $\HI$/${\rm H_{2}}$ decomposition models show qualitatively similar trends. That is, during the quenching of these massive central disc galaxies, as their SFR decreases, their ${\rm H_{2}}$ gas mass and star formation efficiency drop but their $\HI$ gas mass remains about constant. These trends are in general agreement with that reported by \citetalias{Zhang2019}.  
The results from S14 and L08 models match the observations remarkably well. Nevertheless, given the  large systematic uncertainties in the post-processing \citep{Diemer2018}, we avoid stressing the performance of any individual model, but show the variation of all five models in the last panel in Figure~\ref{fig_mhi_sfr_disc}. 
It is important to notice that the $\H2$ mass--SFR relation is more affected by the choices of aperture size. If the SFR($<2R_\star$) is used, in some $\HI/\H2$ decomposition models, the $\H2$ gas stays roughly unchanged with decreasing SFR. This indicates the existence of a population of galaxies with a good amount of SF outside $2R_\star$ in IllustrisTNG. The observed aperture corrected $H_{\alpha}/D_{\rm 4000}$ SFR and UV+IR SFR represent more the SFR over the whole galaxies including the outer region, thus we take SFR(all grav.) as our default SFR choice.
The relations of $\HI$ and $\H2$ mass with SFR agree with the observations within model uncertainties.
We further test our results with SFRs calculated using varying apertures and time-scales, which is shown in Appendix~\ref{sec_SFR}.

We note that for galaxies with SFR below the resolution limit, their median $\HI$ gas mass is still large, with median $M_{\HI}$ larger than $10^{8.6}\,\msun$ when SFR($<2R_\star$) is used and larger than $10^{9.6}\,\msun$ when SFR(all grav.) is used, but in general is lower than those with SFR above the limit in all five models. As shown in the figure, observations do not extend to such low SFR. Hence the existence of these fully quenched massive central disc galaxies with a good amount but lower $\HI$ gas mass becomes a prediction that can be observationally tested by future deeper $\HI$ surveys. 

Although morphology (i.e. disk galaxies) is measured in a different way in the simulations compared to observations.
In Appendix~\ref{sec_morph}, we show that the cold gas--SFR relation in TNG100 and EAGLE is almost independent of morphology, implying that the morphology parameter is irrelevant in interpreting the trends in Figures~\ref{fig_mnh_sfr} and \ref{fig_mhi_sfr_disc} when compared with observations.  However, the independence (or weak dependence) of the $\HI$ mass on morphology clearly contradicts observations. Using the same sample selection, \citetalias{Zhang2019} shows that massive central disc galaxies have an average $\HI$ detection fraction of $>90\%$, while the $\HI$ detection fraction of massive central elliptical galaxies is only $<20\%$.
AGN feedback by itself does not necessarily directly change the morphology. Therefore, the disagreement between observation and simulation may be due to the gas having not been properly processed during morphological transformation in the simulations, such as when mergers and interactions occur.

Star formation efficiency $\epsilon\equiv {\rm SFR}/M_{\rm gas}$ is a widely used observable quantity to explore star formation and quenching in galaxies.
$\epsilon$ (of the total neutral hydrogen gas) in EAGLE and Illustris is approximately constant at $0.5\, {\rm Gyr}^{-1}$, meaning quenching in massive central galaxies in EAGLE and Illustris is primarily driven by the decrease of cold gas. Similarly, in L-Galaxies, $\epsilon$ is approximately constant for galaxies with SFR$>10^{-0.5\,}\msun\, {\rm yr^{-1}}$ and then drops rapidly for the most quenched galaxies.
Hence quenching in L-Galaxies is first driven by the decrease of the cold gas (with a constant $\epsilon$), and then followed by the decrease of $\epsilon$. 
In TNG100, as SFR drops by $\sim 3$ dex, total neutral hydrogen gas mass drops by $\sim 0.5$ dex and hence $\epsilon$ drops by $\sim 2.5$ dex. Therefore, quenching in TNG100 is mainly driven by a decrease in the SF efficiency of the total neutral hydrogen gas. It should be noted that this conclusion relates to the total neutral hydrogen gas, and hence does not contradict the result regarding ${\rm H_2}$ gas shown in Figure~\ref{fig_mhi_sfr_disc} (and also in \citealt{Dou2021a,Dou2021b}) that quenching is driven by the decrease of both ${\rm H_2}$ gas mass and ${\rm H_2}$ star formation efficiency $\epsilon_{\rm H_2}$. It is straightforward to show that 
\begin{equation}
    \epsilon_{\HI+\H2}=\epsilon_{\H2}/(1+M_{\HI}/M_{\H2}).
\end{equation}
The decrease of $\epsilon_{\H2}$ is smaller than that of $\epsilon_{\HI+\H2}$, and the exact amount depends on $M_{\HI}/M_{\H2}$. 
Meanwhile, we stress that in TNG100 the instantaneous SFR in the modeling is calculated based on the local gas density, not based on the ${\H2}$ gas density (which is calculated in post-processing). The dramatic decrease of the $\epsilon_{\HI+\H2}$ of $\sim 2.5$ dex during quenching in TNG100 is caused by the decrease of the gas density likely as a consequence of black hole feedback. Note that in TNG100, when the gas density falls below the minimum threshold (i.e. $0.13\ {\rm cm^{-3}}$), star formation halts, by construction in the simulation. 

We suspect that the large differences in the cold gas content and $\epsilon$ during quenching are mainly caused by the different black hole feedback models implemented in IllustrisTNG (kinetic winds), EAGLE (thermal feedback), Illustris (thermal bubble model) and L-Galaxies (preventive feedback). In many semi-analytic models, such as L-Galaxies, when the average heating rate from BH feedback exceeds the gas cooling rate in the halo, cold gas accretion stops and star formation is quenched. The decrease of the cold gas amount during quenching in central discs in L-Galaxies is caused by the heating from the BH, which reduces the cooling rate of the gas.
In EAGLE, the BH feedback heats up the ambient gas by a temperature increment of $10^{8.5}$\,K; this thermal injection creates a pressure gradient that can produce an outflow. However, this feedback mode might be overly aggressive in heat up cold gas \citep{Dave2020,Mitchell2020}, resulting in a deficiency of the neutral hydrogen gas.
The overall energy injected by BH feedback is similar in IllustrisTNG and Illustris, but the implementation is very different \citep{Pillepich2018a}. 
The thermal bubble model \citep{2007MNRAS.380..877S} in Illustris expels very large amounts of gas from the halo in massive galaxies. The improved kinetic winds \citep{Weinberger2017} in IllustrisTNG still remove large quantities of gas from galaxies during quenching \citep[see related discussions in][]{Stevens2019a,Stevens2021,Terrazas2020}, but it retains more gas on average, as shown in Figure \ref{fig_mnh_sfr} and Figure \ref{fig_sigma_mhi}. We will discuss the effect of BH feedback in IllustrisTNG on gas properties more later.

\subsubsection{Gas distribution profiles}

\begin{figure*}
 \includegraphics[width=0.49\linewidth]{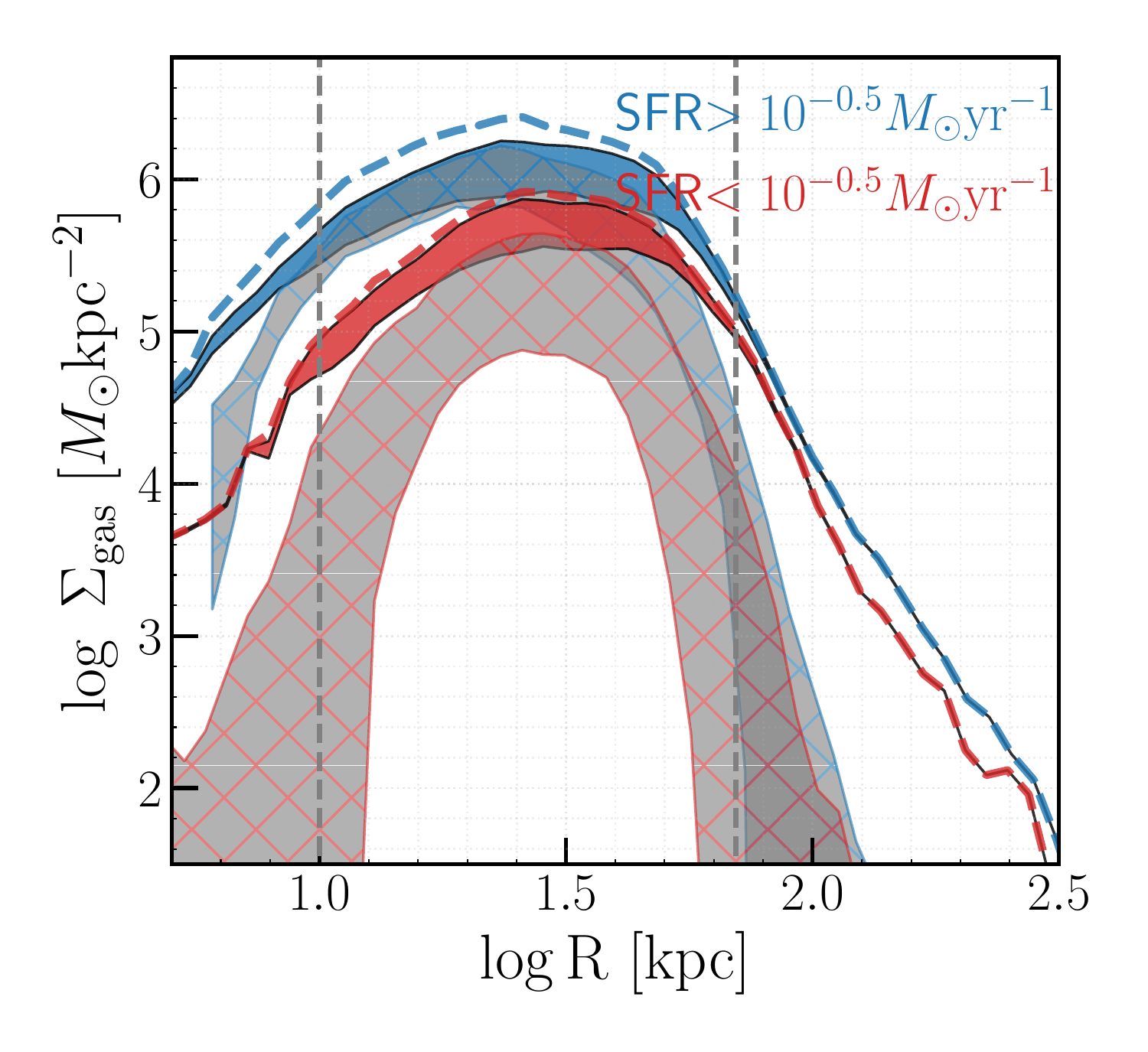}
\centering
 \includegraphics[width=0.49\linewidth]{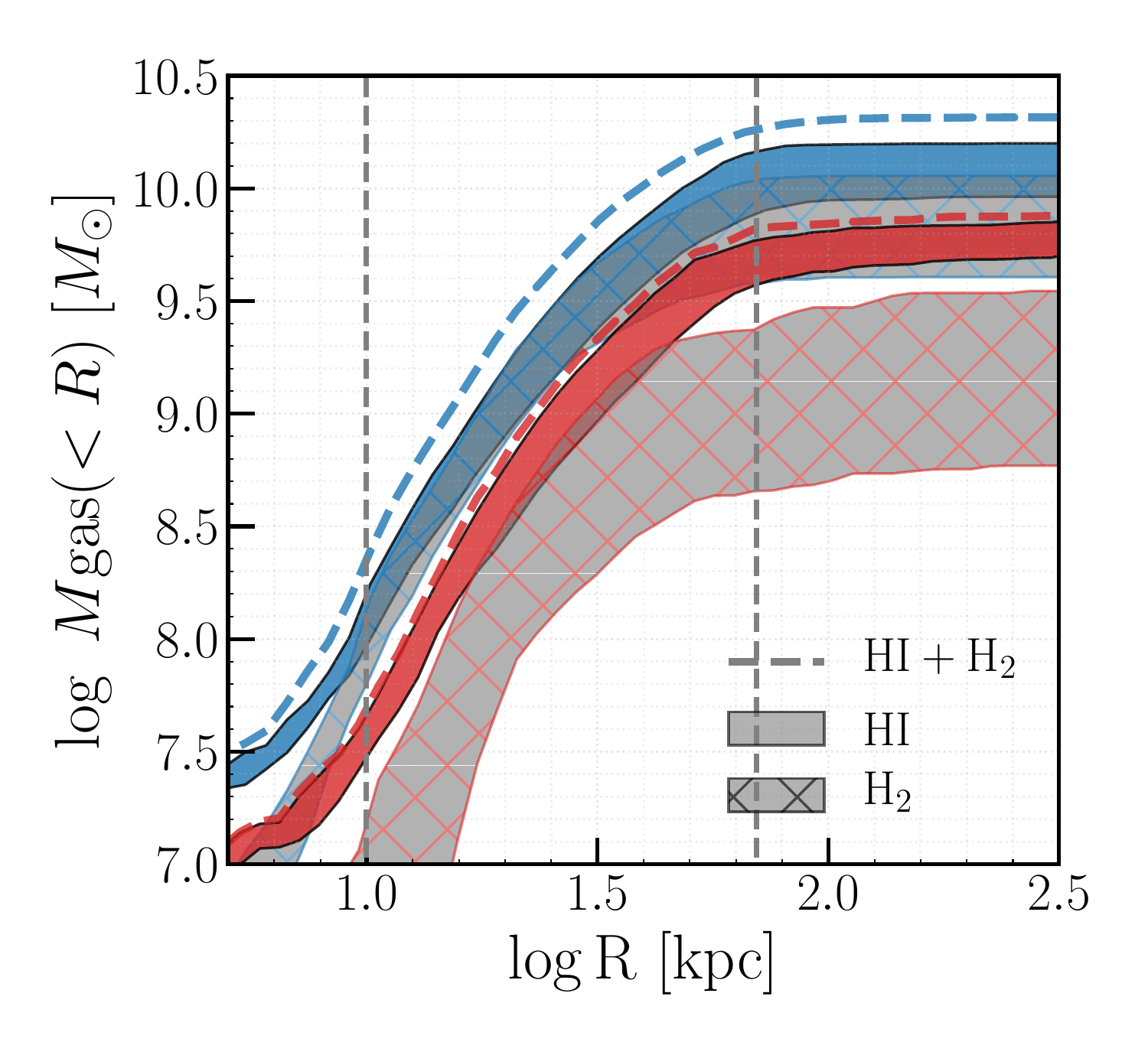}
\caption{\label{fig_sigma_mhi} 
The surface density (face-on) and cumulative radial mass distribution of neutral hydrogen (dashed lines), $\HI$ (shaded area), and $\H2$ (netted shaded region) for higher SFR (blue) and lower SFR (red) central disc galaxies in TNG100, as indicated by the label. The shaded regions indicate the range of median profiles of five $\HI$/${\rm H_2}$ decomposition models. The vertical gray dashed lines correspond to radii of $10$ kpc and $70$ kpc, and are plotted to guide the eye.}
\end{figure*}

Given the fact that TNG100 approximately reproduces the average cold gas mass--SFR relation as shown in Figures \ref{fig_mnh_sfr} and \ref{fig_mhi_sfr_disc}, we further explore in TNG100 the cold gas distributions in galaxies with higher SFR(all grav.)$>10^{-0.5}\msun {\rm yr}^{-1}$ and lower SFR$_{\rm limit}<$SFR(all grav.)$<10^{-0.5}\msun {\rm yr}^{-1}$, to understand the cause of these relations. In the analysis here we exclude the galaxies with the SFR below the resolution limit. 
Current observations in cold gas do not extend to such low SFRs, hence the gas properties in this regime remain a prediction from IllustrisTNG to be tested by future deeper surveys.

Figure \ref{fig_sigma_mhi} shows the median surface density profile and the median cumulative radial distributions of the total neutral hydrogen gas (${\HI+\H2}$), ${\HI}$, and ${\H2}$ for galaxies with higher and lower SFRs in TNG100. The two vertical dashed lines mark radii of $10$ kpc and $70$ kpc to guide the eye. Comparing the total neutral hydrogen gas distribution of galaxies with higher and lower SFR, the difference between the blue and red dashed line is $\sim 0.9$ dex ($\sim 0.7$ dex) in the inner regions at $\sim 10$\,kpc, and is $0.2$ dex ($\sim 0.5$ dex) in the outer regions at $\sim 70$\,kpc for the surface density and cumulative mass. The $\H2$ surface density profile and cumulative mass profiles of galaxies with SFR(all grav.)$>10^{-0.5}\msun {\rm yr}^{-1}$ is similar to that in $\HI$, in particular in the inner regions within $30$\,kpc. By contrast, the ${\H2}$ gas mass of galaxies with SFR(all grav.)$>10^{-0.5}\msun {\rm yr}^{-1}$ is significantly lower than their $\HI$ at all radii. The average $\HI$ gas within $70$\,kpc is $\sim 10^{10} {\rm M}_\odot$ for lower SFR galaxies and $\sim 10^{9.65}\,{\rm M}_\odot$ for higher FSR galaxies. This difference ($\sim 0.35$ dex) is smaller than the difference in the total neutral hydrogen gas, which is $\sim 0.5$ dex at this radius. This makes sense in light of the flat $M_{\HI}$--SFR relation shown in Figure \ref{fig_mhi_sfr_disc}. 

The significant decrease of total neutral hydrogen in the central regions of lower SFR galaxies compared to that of the higher SFR galaxies is likely due to the BH-driven kinetic feedback in IllustrisTNG \citep{Terrazas2020}.
The kinetic winds in IllustrisTNG are implemented by giving the gas immediately surrounding the black hole a momentum kick in a random direction away from the black hole \citep{Weinberger2017}. The kinetic wind this generates is isotropic at the injection scales; it impacts the gas not only in the bi-conical regions, but also in the disc. This kinetic wind quenches galaxy SF by both pushing gas out of galaxies (i.e. ejective) and also heating up the gas (i.e. preventive), more so in the central regions \citep{Zinger2020,Terrazas2020,Truong2020}. Since $\H2$ is mainly located in galaxies' centers, it is the most affected gas phase. While central $\HI$ is also decreased, the majority of $\HI$---which lies in the outskirts of galaxies---is less affected. The deficiency of cold gas in the central regions of the lower SFR galaxies is in agreement with \citetalias{Zhang2019}'s observations that the central regions of the disc galaxies with lower SFR have ${\rm H\alpha/H\beta}$ close to the intrinsic value of $3.1$, which indicates that there is little dust and gas there. Hence Figure~\ref{fig_sigma_mhi} is in fact an important prediction from TNG100, which can be observationally tested with spatially-resolved $\HI$ and $\H2$ observations for both star-forming galaxies and those in the process of being quenched.

Interestingly, at any given radius, the $\H2$ gas mass difference between lower SFR and higher SFR galaxies is significantly larger than their $\HI$ gas mass difference, indicating that the quenching mechanism, i.e. BH feedback, affects $\H2$ more (at the same radius). One important result here is that even in the inner regions of most lowest SFR galaxies, there is a lower but still significant amount of cold gas. 
Therefore, the quenching (i.e. decreasing of SFR) is mainly caused by the fact that most of this neutral hydrogen gas is in the form of non-star-forming $\HI$, not ${\H2}$, which leads to the very low $\epsilon$ of the total neutral hydrogen gas as discussed above.  
As shown in \cite{Diemer2018}, for all of the five $\HI/\H2$ decomposition models, the fraction of neutral hydrogen that is ${\H2}$ depends steeply on the gas density in a certain range (see also \citealt{Morselli2020}). The removal of the high-density gas due to feedback can have a stronger effect on $\H2$ than $\HI$, leading to the low $\H2$-to-$\HI$ ratio, as shown here in Figure \ref{fig_sigma_mhi} and also in Figure \ref{fig_mhi_sfr_disc}. In Appendix~\ref{sec_gas_dens}, we show the stellar light and total gas distribution for two typical central disc galaxies in TNG100, one with SFR(all grav.)$>10^{-0.5}\msun {\rm yr}^{-1}$ and one with SFR(all grav.)$<10^{-0.5}\msun {\rm yr}^{-1}$, to give a more direct view on the above discussion.

%%%%%%%%%%%
%Discussion
%%%%%%%%%%%
\section{Discussion and Summary}
\label{sec_con}

Recent observations indicate the ubiquitous existence of large regularly-rotating $\HI$ discs in massive central disc galaxies, both in those that are star-forming and in those in the process of being quenched \citepalias{Zhang2019}.
This  result has yet been used to calibrate the recipes in the hydrodynamical simulations and semi-analytic models, and can therefore serve as a new observational test of these simulations, in particular of their AGN feedback models, as the existence of an $\HI$ disc at high stellar masses requires the strength and modes of AGN feedback to be just right. If feedback is too violent, $\HI$ discs will not survive during quenching; if it is too weak, it will not be able to deplete the $\H2$ gas in the galaxy and to quench the star formation.

Among the simulations we tested, only TNG100 of the IllustrisTNG project appears to approximately reproduce the trend in cold gas vs SFR as reported in \citetalias{Zhang2019} and some $\H2/\HI$ decomposition method may even reproduce the observed $\H2$ and $\HI$ trends, separately. This lends some support to the AGN feedback implementation in TNG100, though other factors may also be at play. 
Given the coarse resolution of cosmological hdyro-simulation relative to the high-resolution simulation of individual galaxy that can resolve Bondi radius \citep{2015ApJ...804..101Y,2018ApJ...857..121Y} and the sub-grid nature of the the AGN feedback models implemented, it is difficult to tell too much details of the AGN feeback physics. However, reproducing the observed gas content of galaxies of various properties will be critical for future development in cosmological simulations.
Also, the fact that, none of the models tested in this work reproduce the morphology--HI gas mass relation challenges the current models.

In IllustrisTNG, the kinetic winds drive outflows that push gas out of galaxies \citep{Nelson2019b} and increase the gas entropy \citep{Zinger2020}, more so in the central regions, while the $\HI$ gas at larger radii ($\gtrsim 30$kpc) is less affected. This explains the relatively flat $M_{\HI}$--SFR relation shown in Figure \ref{fig_mhi_sfr_disc}.  Meanwhile, the kinetic feedback also leads to a sharp decrease in gas density, which then leads to a sharp decrease in the fraction of molecular hydrogen computed in the decomposition models \citep[cf.][]{Stevens2021}. This explains the rapid decrease of ${\rm H_2}$ gas mass and SFR during quenching. The simultaneous role of being ejective and preventive for AGN feedback in IllustrisTNG also implies that cold gas accretion during quenching is strangulated.
These results are also in qualitative agreement with those derived from the observed SFR/$M_\star$--$M_{\H2}/M_\star$--SFR/$M_{\H2}$ scaling relations \citep{Dou2021a}, in that to quench massive galaxies, strangulation plus additional $\H2$ gas removal (with a mass-loading factor of about unity) are required \citep{Dou2021b}. 

It should be noted that the existence of the observed regularly rotating $\HI$ disk around the massive central disk galaxies that are undergoing quenching process as in  \citetalias{Zhang2019} \&  \citetalias{Zhang2021}, can be explained by AGN feedback as above. Alternatively, it can be caused by angular momentum quenching as proposed in \citet{2020MNRAS.491L..51P} and \citet{2020MNRAS.495L..42R} that the inflowing gas with excess angular momentum can settle on a stable outer neutral hydrogen disk for a long timescale in the absence of perturbation. This is supported by recent studies in simulations which found that a sufficiently high CGM angular momentum is an important factor in keeping a galaxy quenched \citep{2021MNRAS.tmp.2894L,2022MNRAS.509.3148W}. Although AGN feedback must be important to quench massive galaxies in simulations (e.g. \citealt{2019MNRAS.487.4393S}) and can well reproduce the observed cold gas content during quenching as shown in this paper, it may not be the case in the real Universe. We will further investigate both AGN and non-AGN solutions to the quenching of massive galaxies in our future work.

\section{Acknowledgments}
We gratefully acknowledge the anonymous referee for comments and criticisms that have improved the paper.

J.S., Y.P. and L.C.H. acknowledge National Science Foundation of China (NSFC) Grant No. 12125301, 11773001, 12192222, 11721303, 11991052, and the National Key R\&D Program of China grant 2016YFA0400702.

J.S.~acknowledges Martina~Donnari for providing the SFRs of galaxies in TNG100.

A.R.H.S.~acknowledges receipt of the Jim Buckee Fellowship at ICRAR-UWA.

A.R.~acknowledges support from the INAF/PRIN-SKA 2017 `ESKAPEHI' grant.

We acknowledge the science research grants from the China Manned Space Project with NO. CMS-CSST-2021-A04 and NO. CMS-CSST-2021-A07.

Y.G.’s work is partially supported by National Key Basic Research and Development Program of China (grant No. 2017YFA0402704), National Natural Science Foundation of China (NSFC, Nos. 12033004, and 11861131007), and Chinese Academy of Sciences Key Research Program of Frontier Sciences (grant No. QYZDJ-SSW-SLH008). 

%Data 
We acknowledge the Virgo Consortium for making their simulation data available. The eagle simulations were performed using the DiRAC-2 facility at Durham, managed by the ICC, and the PRACE facility Curie based in France at TGCC, CEA, Bruy\'eres-le-Ch\^atel.

\bibliographystyle{mn2e_new}
\bibliography{ref}

\begin{thebibliography}{102}
\expandafter\ifx\csname natexlab\endcsname\relax\def\natexlab#1{#1}\fi

\bibitem[{{Bah{\'e}} {et~al}\mbox{.}(2016){Bah{\'e}}, {Crain}, {Kauffmann},
  {Bower}, {Schaye}, {Furlong}, {Lagos}, {Schaller}, {Trayford}, {Dalla
  Vecchia}, \& {Theuns}}]{Bahe2016}
{Bah{\'e}} Y.~M. {et~al.}, 2016, \mnras, 456, 1115

\bibitem[{{Behroozi} {et~al}\mbox{.}(2019){Behroozi}, {Wechsler}, {Hearin}, \&
  {Conroy}}]{Behroozi2019}
{Behroozi} P., {Wechsler} R.~H., {Hearin} A.~P., {Conroy} C., 2019, \mnras,
  488, 3143

\bibitem[{{Benson} {et~al}\mbox{.}(2003){Benson}, {Bower}, {Frenk}, {Lacey},
  {Baugh}, \& {Cole}}]{Benson2003}
{Benson} A.~J., {Bower} R.~G., {Frenk} C.~S., {Lacey} C.~G., {Baugh} C.~M.,
  {Cole} S., 2003, \apj, 599, 38

\bibitem[{{Bower} {et~al}\mbox{.}(2006){Bower}, {Benson}, {Malbon}, {Helly},
  {Frenk}, {Baugh}, {Cole}, \& {Lacey}}]{Bower2006}
{Bower} R.~G., {Benson} A.~J., {Malbon} R., {Helly} J.~C., {Frenk} C.~S.,
  {Baugh} C.~M., {Cole} S., {Lacey} C.~G., 2006, \mnras, 370, 645

\bibitem[{{Brinchmann} {et~al}\mbox{.}(2004){Brinchmann}, {Charlot}, {White},
  {Tremonti}, {Kauffmann}, {Heckman}, \& {Brinkmann}}]{Brinchmann2004}
{Brinchmann} J., {Charlot} S., {White} S.~D.~M., {Tremonti} C., {Kauffmann} G.,
  {Heckman} T., {Brinkmann} J., 2004, \mnras, 351, 1151

\bibitem[{{Calette} {et~al}\mbox{.}(2018){Calette}, {Avila-Reese},
  {Rodr{\'\i}guez-Puebla}, {Hern{\'a}ndez-Toledo}, \&
  {Papastergis}}]{Calette2018}
{Calette} A.~R., {Avila-Reese} V., {Rodr{\'\i}guez-Puebla} A.,
  {Hern{\'a}ndez-Toledo} H., {Papastergis} E., 2018, \rmxaa, 54, 443

\bibitem[{{Catinella} {et~al}\mbox{.}(2018){Catinella}, {Saintonge},
  {Janowiecki}, {Cortese}, {Dav{\'e}}, {Lemonias}, {Cooper}, {Schiminovich},
  {Hummels}, {Fabello}, {Ger{\'e}b}, {Kilborn}, \& {Wang}}]{Catinella2018}
{Catinella} B. {et~al.}, 2018, \mnras, 476, 875

\bibitem[{{Cortese} {et~al}\mbox{.}(2020){Cortese}, {Catinella}, {Cook}, \&
  {Janowiecki}}]{Cortese2020}
{Cortese} L., {Catinella} B., {Cook} R.~H.~W., {Janowiecki} S., 2020, \mnras,
  494, L42

\bibitem[{{Crain} {et~al}\mbox{.}(2017){Crain}, {Bah{\'e}}, {Lagos}, {Rahmati},
  {Schaye}, {McCarthy}, {Marasco}, {Bower}, {Schaller}, {Theuns}, \& {van der
  Hulst}}]{Crain2017}
{Crain} R.~A. {et~al.}, 2017, \mnras, 464, 4204

\bibitem[{{Crain} {et~al}\mbox{.}(2015){Crain}, {Schaye}, {Bower}, {Furlong},
  {Schaller}, {Theuns}, {Dalla Vecchia}, {Frenk}, {McCarthy}, {Helly},
  {Jenkins}, {Rosas-Guevara}, {White}, \& {Trayford}}]{Crain2015}
---, 2015, \mnras, 450, 1937

\bibitem[{{Cresci} {et~al}\mbox{.}(2015){Cresci}, {Mainieri}, {Brusa},
  {Marconi}, {Perna}, {Mannucci}, {Piconcelli}, {Maiolino}, {Feruglio},
  {Fiore}, {Bongiorno}, {Lanzuisi}, {Merloni}, {Schramm}, {Silverman}, \&
  {Civano}}]{Cresci2015}
{Cresci} G. {et~al.}, 2015, \apj, 799, 82

\bibitem[{{Cresci} \& {Maiolino}(2018)}]{Cresci2018}
{Cresci} G., {Maiolino} R., 2018, Nature Astronomy, 2, 179

\bibitem[{{Croton} {et~al}\mbox{.}(2006){Croton}, {Springel}, {White}, {De
  Lucia}, {Frenk}, {Gao}, {Jenkins}, {Kauffmann}, {Navarro}, \&
  {Yoshida}}]{Croton2006}
{Croton} D.~J. {et~al.}, 2006, \mnras, 365, 11

\bibitem[{{Dav{\'e}} {et~al}\mbox{.}(2019){Dav{\'e}}, {Angl{\'e}s-Alc{\'a}zar},
  {Narayanan}, {Li}, {Rafieferantsoa}, \& {Appleby}}]{Dave2019}
{Dav{\'e}} R., {Angl{\'e}s-Alc{\'a}zar} D., {Narayanan} D., {Li} Q.,
  {Rafieferantsoa} M.~H., {Appleby} S., 2019, \mnras, 486, 2827

\bibitem[{{Dav{\'e}} {et~al}\mbox{.}(2020){Dav{\'e}}, {Crain}, {Stevens},
  {Narayanan}, {Saintonge}, {Catinella}, \& {Cortese}}]{Dave2020}
{Dav{\'e}} R., {Crain} R.~A., {Stevens} A. R.~H., {Narayanan} D., {Saintonge}
  A., {Catinella} B., {Cortese} L., 2020, \mnras, 497, 146

\bibitem[{{Davis} {et~al}\mbox{.}(1985){Davis}, {Efstathiou}, {Frenk}, \&
  {White}}]{Davis1985}
{Davis} M., {Efstathiou} G., {Frenk} C.~S., {White} S.~D.~M., 1985, \apj, 292,
  371

\bibitem[{{Diemer} {et~al}\mbox{.}(2018){Diemer}, {Stevens}, {Forbes},
  {Marinacci}, {Hernquist}, {Lagos}, {Sternberg}, {Pillepich}, {Nelson},
  {Popping}, {Villaescusa-Navarro}, {Torrey}, \& {Vogelsberger}}]{Diemer2018}
{Diemer} B. {et~al.}, 2018, \apjs, 238, 33

\bibitem[{{Diemer} {et~al}\mbox{.}(2019){Diemer}, {Stevens}, {Lagos},
  {Calette}, {Tacchella}, {Hernquist}, {Marinacci}, {Nelson}, {Pillepich},
  {Rodriguez-Gomez}, {Villaescusa-Navarro}, \& {Vogelsberger}}]{Diemer2019}
---, 2019, \mnras, 487, 1529

\bibitem[{{Donnari} {et~al}\mbox{.}(2019){Donnari}, {Pillepich}, {Nelson},
  {Vogelsberger}, {Genel}, {Weinberger}, {Marinacci}, {Springel}, \&
  {Hernquist}}]{Donnari2019}
{Donnari} M. {et~al.}, 2019, \mnras, 485, 4817

\bibitem[{{Dou} {et~al}\mbox{.}(2021{\natexlab{a}}){Dou}, {Peng}, {Renzini},
  {Ho}, {Mannucci}, {Daddi}, {Gao}, {Maiolino}, {Zhang}, {Gu}, {Li}, {Lilly},
  {Pan}, {Yuan}, \& {Zheng}}]{Dou2021b}
{Dou} J. {et~al.}, 2021{\natexlab{a}}, arXiv e-prints, arXiv:2104.12794

\bibitem[{{Dou} {et~al}\mbox{.}(2021{\natexlab{b}}){Dou}, {Peng}, {Renzini},
  {Ho}, {Mannucci}, {Daddi}, {Gao}, {Maiolino}, {Zhang}, {Gu}, {Li}, {Lilly},
  \& {Yuan}}]{Dou2021a}
---, 2021{\natexlab{b}}, \apj, 907, 114

\bibitem[{{Dutton} {et~al}\mbox{.}(2010){Dutton}, {Conroy}, {van den Bosch},
  {Prada}, \& {More}}]{Dutton2010}
{Dutton} A.~A., {Conroy} C., {van den Bosch} F.~C., {Prada} F., {More} S.,
  2010, \mnras, 407, 2

\bibitem[{{Gallagher} {et~al}\mbox{.}(2019){Gallagher}, {Maiolino}, {Belfiore},
  {Drory}, {Riffel}, \& {Riffel}}]{Gallagher2019}
{Gallagher} R., {Maiolino} R., {Belfiore} F., {Drory} N., {Riffel} R., {Riffel}
  R.~A., 2019, \mnras, 485, 3409

\bibitem[{{Genel} {et~al}\mbox{.}(2014){Genel}, {Vogelsberger}, {Springel},
  {Sijacki}, {Nelson}, {Snyder}, {Rodriguez-Gomez}, {Torrey}, \&
  {Hernquist}}]{Genel2014}
{Genel} S. {et~al.}, 2014, \mnras, 445, 175

\bibitem[{{Gnedin} \& {Draine}(2014)}]{Gnedin2014}
{Gnedin} N.~Y., {Draine} B.~T., 2014, \apj, 795, 37

\bibitem[{{Gnedin} \& {Kravtsov}(2011)}]{Gnedin2011}
{Gnedin} N.~Y., {Kravtsov} A.~V., 2011, \apj, 728, 88

\bibitem[{{Guo} {et~al}\mbox{.}(2011){Guo}, {White}, {Boylan-Kolchin}, {De
  Lucia}, {Kauffmann}, {Lemson}, {Li}, {Springel}, \& {Weinmann}}]{Guo2011}
{Guo} Q. {et~al.}, 2011, \mnras, 413, 101

\bibitem[{{Henriques} {et~al}\mbox{.}(2015){Henriques}, {White}, {Thomas},
  {Angulo}, {Guo}, {Lemson}, {Springel}, \& {Overzier}}]{Henriques2015}
{Henriques} B. M.~B., {White} S. D.~M., {Thomas} P.~A., {Angulo} R., {Guo} Q.,
  {Lemson} G., {Springel} V., {Overzier} R., 2015, \mnras, 451, 2663

\bibitem[{{Henriques} {et~al}\mbox{.}(2020){Henriques}, {Yates}, {Fu}, {Guo},
  {Kauffmann}, {Srisawat}, {Thomas}, \& {White}}]{Henriques2020}
{Henriques} B. M.~B., {Yates} R.~M., {Fu} J., {Guo} Q., {Kauffmann} G.,
  {Srisawat} C., {Thomas} P.~A., {White} S. D.~M., 2020, \mnras, 491, 5795

\bibitem[{{Ishibashi} \& {Fabian}(2012)}]{Ishibashi2012}
{Ishibashi} W., {Fabian} A.~C., 2012, \mnras, 427, 2998

\bibitem[{{Jones} {et~al}\mbox{.}(2018){Jones}, {Haynes}, {Giovanelli}, \&
  {Moorman}}]{Jones2018}
{Jones} M.~G., {Haynes} M.~P., {Giovanelli} R., {Moorman} C., 2018, \mnras,
  477, 2

\bibitem[{{Kravtsov}, {Vikhlinin} \& {Meshcheryakov}(2018){Kravtsov},
  {Vikhlinin}, \& {Meshcheryakov}}]{Kravtsov2018}
{Kravtsov} A.~V., {Vikhlinin} A.~A., {Meshcheryakov} A.~V., 2018, Astronomy
  Letters, 44, 8

\bibitem[{{Krumholz}(2013)}]{Krumholz2013}
{Krumholz} M.~R., 2013, \mnras, 436, 2747

\bibitem[{{Lagos} {et~al}\mbox{.}(2015){Lagos}, {Crain}, {Schaye}, {Furlong},
  {Frenk}, {Bower}, {Schaller}, {Theuns}, {Trayford}, {Bah{\'e}}, \& {Dalla
  Vecchia}}]{Lagos2015}
{Lagos} C. d.~P. {et~al.}, 2015, \mnras, 452, 3815

\bibitem[{{Lagos} {et~al}\mbox{.}(2018){Lagos}, {Tobar}, {Robotham},
  {Obreschkow}, {Mitchell}, {Power}, \& {Elahi}}]{2018MNRAS.481.3573L}
{Lagos} C. d.~P., {Tobar} R.~J., {Robotham} A. S.~G., {Obreschkow} D.,
  {Mitchell} P.~D., {Power} C., {Elahi} P.~J., 2018, \mnras, 481, 3573

\bibitem[{{Larson}(1974)}]{Larson1974}
{Larson} R.~B., 1974, \mnras, 169, 229

\bibitem[{{Leroy} {et~al}\mbox{.}(2008){Leroy}, {Walter}, {Brinks}, {Bigiel},
  {de Blok}, {Madore}, \& {Thornley}}]{Leroy2008}
{Leroy} A.~K., {Walter} F., {Brinks} E., {Bigiel} F., {de Blok} W.~J.~G.,
  {Madore} B., {Thornley} M.~D., 2008, \aj, 136, 2782

\bibitem[{{Lu} {et~al}\mbox{.}(2021){Lu}, {Xu}, {Wang}, {Cai}, {He}, {Xu},
  {Xia}, {Mao}, {Springel}, \& {Hernquist}}]{2021MNRAS.tmp.2894L}
{Lu} S. {et~al.}, 2021, \mnras

\bibitem[{{Maiolino} {et~al}\mbox{.}(2017){Maiolino}, {Russell}, {Fabian},
  {Carniani}, {Gallagher}, {Cazzoli}, {Arribas}, {Belfiore}, {Bellocchi},
  {Colina}, {Cresci}, {Ishibashi}, {Marconi}, {Mannucci}, {Oliva}, \&
  {Sturm}}]{Maiolino2017}
{Maiolino} R. {et~al.}, 2017, \nat, 544, 202

\bibitem[{{Marinacci} {et~al}\mbox{.}(2018){Marinacci}, {Vogelsberger},
  {Pakmor}, {Torrey}, {Springel}, {Hernquist}, {Nelson}, {Weinberger},
  {Pillepich}, {Naiman}, \& {Genel}}]{Marinacci2018}
{Marinacci} F. {et~al.}, 2018, \mnras, 480, 5113

\bibitem[{{McAlpine} {et~al}\mbox{.}(2016){McAlpine}, {Helly}, {Schaller},
  {Trayford}, {Qu}, {Furlong}, {Bower}, {Crain}, {Schaye}, {Theuns}, {Dalla
  Vecchia}, {Frenk}, {McCarthy}, {Jenkins}, {Rosas-Guevara}, {White}, {Baes},
  {Camps}, \& {Lemson}}]{McAlpine2016}
{McAlpine} S. {et~al.}, 2016, Astronomy and Computing, 15, 72

\bibitem[{{Mitchell} {et~al}\mbox{.}(2020){Mitchell}, {Schaye}, {Bower}, \&
  {Crain}}]{Mitchell2020}
{Mitchell} P.~D., {Schaye} J., {Bower} R.~G., {Crain} R.~A., 2020, \mnras, 494,
  3971

\bibitem[{{Morselli} {et~al}\mbox{.}(2020){Morselli}, {Rodighiero}, {Enia},
  {Corbelli}, {Casasola}, {Rodr{\'\i}guez-Mu{\~n}oz}, {Renzini}, {Tacchella},
  {Baronchelli}, {Bianchi}, {Cassata}, {Franceschini}, {Mancini}, {Negrello},
  {Popesso}, \& {Romano}}]{Morselli2020}
{Morselli} L. {et~al.}, 2020, \mnras, 496, 4606

\bibitem[{{Moster}, {Naab} \& {White}(2018){Moster}, {Naab}, \&
  {White}}]{Moster2018}
{Moster} B.~P., {Naab} T., {White} S. D.~M., 2018, \mnras, 477, 1822

\bibitem[{{Naab} \& {Ostriker}(2017)}]{NaabOstriker2017}
{Naab} T., {Ostriker} J.~P., 2017, \araa, 55, 59

\bibitem[{{Naiman} {et~al}\mbox{.}(2018){Naiman}, {Pillepich}, {Springel},
  {Ramirez-Ruiz}, {Torrey}, {Vogelsberger}, {Pakmor}, {Nelson}, {Marinacci},
  {Hernquist}, {Weinberger}, \& {Genel}}]{Naiman2018}
{Naiman} J.~P. {et~al.}, 2018, \mnras, 477, 1206

\bibitem[{{Nelson} {et~al}\mbox{.}(2015){Nelson}, {Pillepich}, {Genel},
  {Vogelsberger}, {Springel}, {Torrey}, {Rodriguez-Gomez}, {Sijacki}, {Snyder},
  {Griffen}, {Marinacci}, {Blecha}, {Sales}, {Xu}, \& {Hernquist}}]{Nelson2015}
{Nelson} D. {et~al.}, 2015, Astronomy and Computing, 13, 12

\bibitem[{{Nelson} {et~al}\mbox{.}(2019{\natexlab{a}}){Nelson}, {Pillepich},
  {Springel}, {Pakmor}, {Weinberger}, {Genel}, {Torrey}, {Vogelsberger},
  {Marinacci}, \& {Hernquist}}]{Nelson2019b}
---, 2019{\natexlab{a}}, \mnras, 490, 3234

\bibitem[{{Nelson} {et~al}\mbox{.}(2018){Nelson}, {Pillepich}, {Springel},
  {Weinberger}, {Hernquist}, {Pakmor}, {Genel}, {Torrey}, {Vogelsberger},
  {Kauffmann}, {Marinacci}, \& {Naiman}}]{Nelson2018}
---, 2018, \mnras, 475, 624

\bibitem[{{Nelson} {et~al}\mbox{.}(2019{\natexlab{b}}){Nelson}, {Springel},
  {Pillepich}, {Rodriguez-Gomez}, {Torrey}, {Genel}, {Vogelsberger}, {Pakmor},
  {Marinacci}, {Weinberger}, {Kelley}, {Lovell}, {Diemer}, \&
  {Hernquist}}]{Nelson2019}
---, 2019{\natexlab{b}}, Computational Astrophysics and Cosmology, 6, 2

\bibitem[{{Pawlik} \& {Schaye}(2008)}]{Pawlik2008}
{Pawlik} A.~H., {Schaye} J., 2008, \mnras, 389, 651

\bibitem[{{Peng} \& {Renzini}(2020)}]{2020MNRAS.491L..51P}
{Peng} Y.-j., {Renzini} A., 2020, \mnras, 491, L51

\bibitem[{{Pillepich} {et~al}\mbox{.}(2018{\natexlab{a}}){Pillepich}, {Nelson},
  {Hernquist}, {Springel}, {Pakmor}, {Torrey}, {Weinberger}, {Genel}, {Naiman},
  {Marinacci}, \& {Vogelsberger}}]{Pillepich2018b}
{Pillepich} A. {et~al.}, 2018{\natexlab{a}}, \mnras, 475, 648

\bibitem[{{Pillepich} {et~al}\mbox{.}(2018{\natexlab{b}}){Pillepich},
  {Springel}, {Nelson}, {Genel}, {Naiman}, {Pakmor}, {Hernquist}, {Torrey},
  {Vogelsberger}, {Weinberger}, \& {Marinacci}}]{Pillepich2018a}
---, 2018{\natexlab{b}}, \mnras, 473, 4077

\bibitem[{{Rahmati} {et~al}\mbox{.}(2013){Rahmati}, {Pawlik},
  {Rai{\v{c}}evi{\'c}}, \& {Schaye}}]{Rahmati2013}
{Rahmati} A., {Pawlik} A.~H., {Rai{\v{c}}evi{\'c}} M., {Schaye} J., 2013,
  \mnras, 430, 2427

\bibitem[{{Rees} \& {Ostriker}(1977)}]{ReesOstriker1977}
{Rees} M.~J., {Ostriker} J.~P., 1977, \mnras, 179, 541

\bibitem[{{Renzini}(2020)}]{2020MNRAS.495L..42R}
{Renzini} A., 2020, \mnras, 495, L42

\bibitem[{{Rodriguez-Gomez} {et~al}\mbox{.}(2015){Rodriguez-Gomez}, {Genel},
  {Vogelsberger}, {Sijacki}, {Pillepich}, {Sales}, {Torrey}, {Snyder},
  {Nelson}, {Springel}, {Ma}, \& {Hernquist}}]{Rodriguez-Gomez2015}
{Rodriguez-Gomez} V. {et~al.}, 2015, \mnras, 449, 49

\bibitem[{{Saintonge} {et~al}\mbox{.}(2016){Saintonge}, {Catinella}, {Cortese},
  {Genzel}, {Giovanelli}, {Haynes}, {Janowiecki}, {Kramer}, {Lutz},
  {Schiminovich}, {Tacconi}, {Wuyts}, \& {Accurso}}]{Saintonge2016}
{Saintonge} A. {et~al.}, 2016, \mnras, 462, 1749

\bibitem[{{Sales} {et~al}\mbox{.}(2012){Sales}, {Navarro}, {Theuns}, {Schaye},
  {White}, {Frenk}, {Crain}, \& {Dalla Vecchia}}]{Sales2012}
{Sales} L.~V., {Navarro} J.~F., {Theuns} T., {Schaye} J., {White} S. D.~M.,
  {Frenk} C.~S., {Crain} R.~A., {Dalla Vecchia} C., 2012, \mnras, 423, 1544

\bibitem[{{Salim}, {Boquien} \& {Lee}(2018){Salim}, {Boquien}, \&
  {Lee}}]{Salim2018}
{Salim} S., {Boquien} M., {Lee} J.~C., 2018, \apj, 859, 11

\bibitem[{{Salim} {et~al}\mbox{.}(2016){Salim}, {Lee}, {Janowiecki}, {da
  Cunha}, {Dickinson}, {Boquien}, {Burgarella}, {Salzer}, \&
  {Charlot}}]{Salim2016}
{Salim} S. {et~al.}, 2016, \apjs, 227, 2

\bibitem[{{Schaye} {et~al}\mbox{.}(2015){Schaye}, {Crain}, {Bower}, {Furlong},
  {Schaller}, {Theuns}, {Dalla Vecchia}, {Frenk}, {McCarthy}, {Helly},
  {Jenkins}, {Rosas-Guevara}, {White}, {Baes}, {Booth}, {Camps}, {Navarro},
  {Qu}, {Rahmati}, {Sawala}, {Thomas}, \& {Trayford}}]{Schaye2015}
{Schaye} J. {et~al.}, 2015, \mnras, 446, 521

\bibitem[{{Sijacki} {et~al}\mbox{.}(2007){Sijacki}, {Springel}, {Di Matteo}, \&
  {Hernquist}}]{2007MNRAS.380..877S}
{Sijacki} D., {Springel} V., {Di Matteo} T., {Hernquist} L., 2007, \mnras, 380,
  877

\bibitem[{{Sijacki} {et~al}\mbox{.}(2015){Sijacki}, {Vogelsberger}, {Genel},
  {Springel}, {Torrey}, {Snyder}, {Nelson}, \& {Hernquist}}]{Sijacki2015}
{Sijacki} D., {Vogelsberger} M., {Genel} S., {Springel} V., {Torrey} P.,
  {Snyder} G.~F., {Nelson} D., {Hernquist} L., 2015, \mnras, 452, 575

\bibitem[{{Silk}(2003)}]{Silk2003}
{Silk} J., 2003, \mnras, 343, 249

\bibitem[{{Silk}(2013)}]{Silk2013}
---, 2013, \apj, 772, 112

\bibitem[{{Silk} \& {Rees}(1998)}]{Silk1998}
{Silk} J., {Rees} M.~J., 1998, \aap, 331, L1

\bibitem[{{Somerville} \& {Dav{\'e}}(2015)}]{SomervilleDave2015}
{Somerville} R.~S., {Dav{\'e}} R., 2015, \araa, 53, 51

\bibitem[{{Speagle} {et~al}\mbox{.}(2014){Speagle}, {Steinhardt}, {Capak}, \&
  {Silverman}}]{2014ApJS..214...15S}
{Speagle} J.~S., {Steinhardt} C.~L., {Capak} P.~L., {Silverman} J.~D., 2014,
  \apjs, 214, 15

\bibitem[{{Springel}(2010)}]{Springel2010}
{Springel} V., 2010, \mnras, 401, 791

\bibitem[{{Springel} \& {Hernquist}(2003)}]{Springel2003}
{Springel} V., {Hernquist} L., 2003, \mnras, 339, 289

\bibitem[{{Springel} {et~al}\mbox{.}(2018){Springel}, {Pakmor}, {Pillepich},
  {Weinberger}, {Nelson}, {Hernquist}, {Vogelsberger}, {Genel}, {Torrey},
  {Marinacci}, \& {Naiman}}]{Springel2018}
{Springel} V. {et~al.}, 2018, \mnras, 475, 676

\bibitem[{{Springel} {et~al}\mbox{.}(2001){Springel}, {White}, {Tormen}, \&
  {Kauffmann}}]{Springel2001}
{Springel} V., {White} S. D.~M., {Tormen} G., {Kauffmann} G., 2001, \mnras,
  328, 726

\bibitem[{{Sternberg} {et~al}\mbox{.}(2014){Sternberg}, {Le Petit}, {Roueff},
  \& {Le Bourlot}}]{Sternberg2014}
{Sternberg} A., {Le Petit} F., {Roueff} E., {Le Bourlot} J., 2014, \apj, 790,
  10

\bibitem[{{Stevens} {et~al}\mbox{.}(2019{\natexlab{a}}){Stevens}, {Diemer},
  {Lagos}, {Nelson}, {Obreschkow}, {Wang}, \& {Marinacci}}]{Stevens2019b}
{Stevens} A. R.~H., {Diemer} B., {Lagos} C. d.~P., {Nelson} D., {Obreschkow}
  D., {Wang} J., {Marinacci} F., 2019{\natexlab{a}}, \mnras, 490, 96

\bibitem[{{Stevens} {et~al}\mbox{.}(2019{\natexlab{b}}){Stevens}, {Diemer},
  {Lagos}, {Nelson}, {Pillepich}, {Brown}, {Catinella}, {Hernquist},
  {Weinberger}, {Vogelsberger}, \& {Marinacci}}]{Stevens2019a}
{Stevens} A. R.~H. {et~al.}, 2019{\natexlab{b}}, \mnras, 483, 5334

\bibitem[{{Stevens} {et~al}\mbox{.}(2021){Stevens}, {Lagos}, {Cortese},
  {Catinella}, {Diemer}, {Nelson}, {Pillepich}, {Hernquist}, {Marinacci}, \&
  {Vogelsberger}}]{Stevens2021}
---, 2021, \mnras, 502, 3158

\bibitem[{{Su} {et~al}\mbox{.}(2019){Su}, {Hopkins}, {Hayward}, {Ma},
  {Faucher-Gigu{\`e}re}, {Kere{\v{s}}}, {Orr}, {Chan}, \&
  {Robles}}]{2019MNRAS.487.4393S}
{Su} K.-Y. {et~al.}, 2019, \mnras, 487, 4393

\bibitem[{{Tacconi} {et~al}\mbox{.}(2018){Tacconi}, {Genzel}, {Saintonge},
  {Combes}, {Garc{\'\i}a-Burillo}, {Neri}, {Bolatto}, {Contini}, {F{\"o}rster
  Schreiber}, {Lilly}, {Lutz}, {Wuyts}, {Accurso}, {Boissier}, {Boone},
  {Bouch{\'e}}, {Bournaud}, {Burkert}, {Carollo}, {Cooper}, {Cox}, {Feruglio},
  {Freundlich}, {Herrera-Camus}, {Juneau}, {Lippa}, {Naab}, {Renzini},
  {Salome}, {Sternberg}, {Tadaki}, {{\"U}bler}, {Walter}, {Weiner}, \&
  {Weiss}}]{Tacconi2018}
{Tacconi} L.~J. {et~al.}, 2018, \apj, 853, 179

\bibitem[{{Terrazas} {et~al}\mbox{.}(2020){Terrazas}, {Bell}, {Pillepich},
  {Nelson}, {Somerville}, {Genel}, {Weinberger}, {Habouzit}, {Li}, {Hernquist},
  \& {Vogelsberger}}]{Terrazas2020}
{Terrazas} B.~A. {et~al.}, 2020, \mnras, 493, 1888

\bibitem[{{Torrey} {et~al}\mbox{.}(2014){Torrey}, {Vogelsberger}, {Genel},
  {Sijacki}, {Springel}, \& {Hernquist}}]{Torrey2014}
{Torrey} P., {Vogelsberger} M., {Genel} S., {Sijacki} D., {Springel} V.,
  {Hernquist} L., 2014, \mnras, 438, 1985

\bibitem[{{Truong} {et~al}\mbox{.}(2020){Truong}, {Pillepich}, {Werner},
  {Nelson}, {Lakhchaura}, {Weinberger}, {Springel}, {Vogelsberger}, \&
  {Hernquist}}]{Truong2020}
{Truong} N. {et~al.}, 2020, \mnras, 494, 549

\bibitem[{{Vogelsberger} {et~al}\mbox{.}(2013){Vogelsberger}, {Genel},
  {Sijacki}, {Torrey}, {Springel}, \& {Hernquist}}]{Vogelsberger2013}
{Vogelsberger} M., {Genel} S., {Sijacki} D., {Torrey} P., {Springel} V.,
  {Hernquist} L., 2013, \mnras, 436, 3031

\bibitem[{{Vogelsberger} {et~al}\mbox{.}(2014){Vogelsberger}, {Genel},
  {Springel}, {Torrey}, {Sijacki}, {Xu}, {Snyder}, {Bird}, {Nelson}, \&
  {Hernquist}}]{Vogelsberger2014}
{Vogelsberger} M. {et~al.}, 2014, \nat, 509, 177

\bibitem[{{Wang} \& {Jing}(2010)}]{WangJing2010}
{Wang} L., {Jing} Y.~P., 2010, \mnras, 402, 1796

\bibitem[{{Wang} {et~al}\mbox{.}(2022){Wang}, {Xu}, {Lu}, {Cai}, {Xiang},
  {Mao}, {Springel}, \& {Hernquist}}]{2022MNRAS.509.3148W}
{Wang} S., {Xu} D., {Lu} S., {Cai} Z., {Xiang} M., {Mao} S., {Springel} V.,
  {Hernquist} L., 2022, \mnras, 509, 3148

\bibitem[{{Wechsler} \& {Tinker}(2018)}]{WechslerTinker2018}
{Wechsler} R.~H., {Tinker} J.~L., 2018, \araa, 56, 435

\bibitem[{{Weinberger} {et~al}\mbox{.}(2017){Weinberger}, {Springel},
  {Hernquist}, {Pillepich}, {Marinacci}, {Pakmor}, {Nelson}, {Genel},
  {Vogelsberger}, {Naiman}, \& {Torrey}}]{Weinberger2017}
{Weinberger} R. {et~al.}, 2017, \mnras, 465, 3291

\bibitem[{{White} \& {Frenk}(1991)}]{WhiteFrenk1991}
{White} S. D.~M., {Frenk} C.~S., 1991, \apj, 379, 52

\bibitem[{{White} \& {Rees}(1978)}]{WhiteRees1978}
{White} S.~D.~M., {Rees} M.~J., 1978, \mnras, 183, 341

\bibitem[{{Yang}, {Mo} \& {van den Bosch}(2009){Yang}, {Mo}, \& {van den
  Bosch}}]{Yang2009}
{Yang} X., {Mo} H.~J., {van den Bosch} F.~C., 2009, \apj, 695, 900

\bibitem[{{Yang} {et~al}\mbox{.}(2012){Yang}, {Mo}, {van den Bosch}, {Zhang},
  \& {Han}}]{Yang2012}
{Yang} X., {Mo} H.~J., {van den Bosch} F.~C., {Zhang} Y., {Han} J., 2012, \apj,
  752, 41

\bibitem[{{Yates} {et~al}\mbox{.}(2013){Yates}, {Henriques}, {Thomas},
  {Kauffmann}, {Johansson}, \& {White}}]{Yates2013}
{Yates} R.~M., {Henriques} B., {Thomas} P.~A., {Kauffmann} G., {Johansson} J.,
  {White} S. D.~M., 2013, \mnras, 435, 3500

\bibitem[{{Yuan} {et~al}\mbox{.}(2015){Yuan}, {Gan}, {Narayan}, {Sadowski},
  {Bu}, \& {Bai}}]{2015ApJ...804..101Y}
{Yuan} F., {Gan} Z., {Narayan} R., {Sadowski} A., {Bu} D., {Bai} X.-N., 2015,
  \apj, 804, 101

\bibitem[{{Yuan} \& {Narayan}(2014)}]{2014ARA&A..52..529Y}
{Yuan} F., {Narayan} R., 2014, \araa, 52, 529

\bibitem[{{Yuan} {et~al}\mbox{.}(2018){Yuan}, {Yoon}, {Li}, {Gan}, {Ho}, \&
  {Guo}}]{2018ApJ...857..121Y}
{Yuan} F., {Yoon} D., {Li} Y.-P., {Gan} Z.-M., {Ho} L.~C., {Guo} F., 2018,
  \apj, 857, 121

\bibitem[{{Zhang} {et~al}\mbox{.}(2019){Zhang}, {Peng}, {Ho}, {Maiolino},
  {Dekel}, {Guo}, {Mannucci}, {Li}, {Yuan}, {Renzini}, {Dou}, {Guo}, {Man}, \&
  {Li}}]{Zhang2019}
{Zhang} C. {et~al.}, 2019, \apjl, 884, L52

\bibitem[{{Zhang} {et~al}\mbox{.}(2021){Zhang}, {Peng}, {Ho}, {Maiolino},
  {Renzini}, {Mannucci}, {Dekel}, {Guo}, {Li}, {Yuan}, {Lilly}, {Dou}, {Guo},
  {Man}, {Li}, \& {Shi}}]{Zhang2021}
---, 2021, \apj, 911, 57

\bibitem[{{Zheng}, {Coil} \& {Zehavi}(2007){Zheng}, {Coil}, \&
  {Zehavi}}]{Zheng2007}
{Zheng} Z., {Coil} A.~L., {Zehavi} I., 2007, \apj, 667, 760

\bibitem[{{Zinger} {et~al}\mbox{.}(2020){Zinger}, {Pillepich}, {Nelson},
  {Weinberger}, {Pakmor}, {Springel}, {Hernquist}, {Marinacci}, \&
  {Vogelsberger}}]{Zinger2020}
{Zinger} E. {et~al.}, 2020, \mnras, 499, 768

\bibitem[{{Zubovas} {et~al}\mbox{.}(2013){Zubovas}, {Nayakshin}, {King}, \&
  {Wilkinson}}]{Zubovas2013}
{Zubovas} K., {Nayakshin} S., {King} A., {Wilkinson} M., 2013, \mnras, 433,
  3079

\end{thebibliography}

%%%%%%%%%%%
%Appendix
%%%%%%%%%%%
\newpage
\appendix
\counterwithin{figure}{section}

\section{Cold Gas--SFR Relation for Centrals in TNG100 and EAGLE}
\label{sec_morph}

\begin{figure*}
\centering
 \includegraphics[width=1.\linewidth]{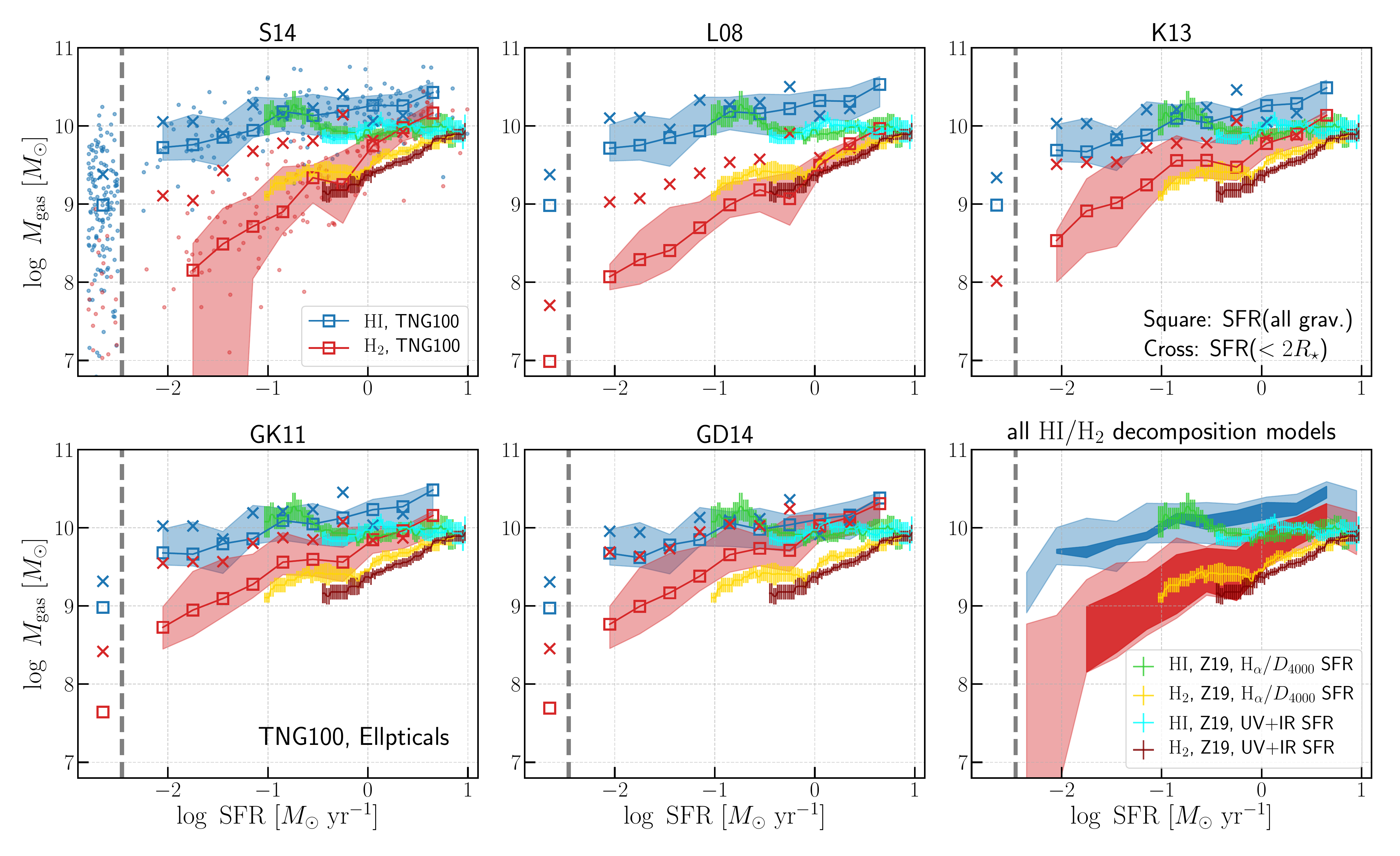}
\caption{\label{fig_mhi_sfr_el} Similar to Figure~\ref{fig_mhi_sfr_disc}, but for ellipticals in TNG100. Note that here the observations are for central discs.}
\end{figure*}

\begin{figure*}
\centering
 \includegraphics[width=.9\linewidth]{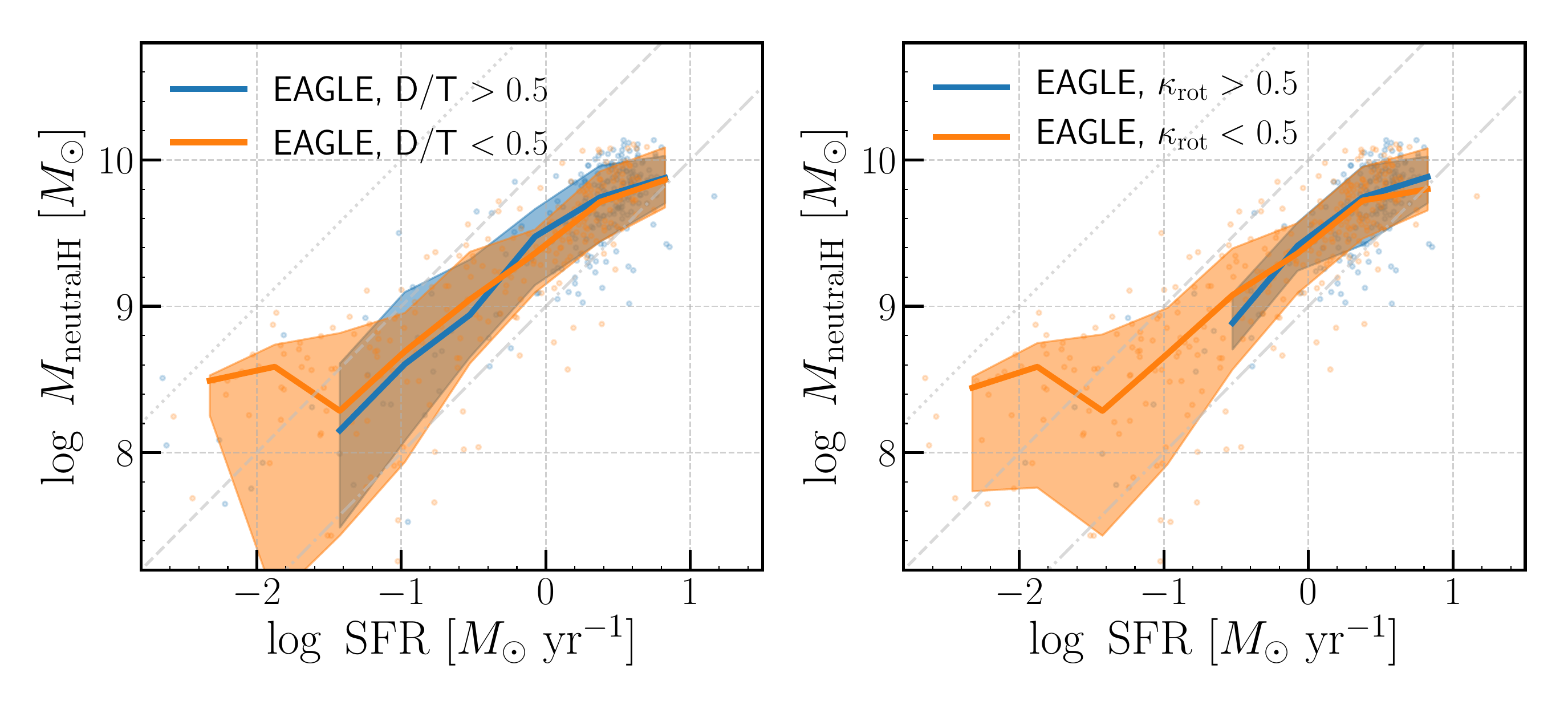}
\caption{\label{fig_mhi_sfr_centrals_eagle} Neutral hydrogen gas mass versus SFR for central discs and ellipticals in EAGLE. The left- and right-hand panels respectively show the morphology divisions using disc-to-total mass ratio (D/T) and rotation support parameter ($\kappa_{\rm rot}$). The blue and oranges lines are median relations for discs and ellipticals separately, together with the 16th--84th percentile ranges shown by the shaded areas.}
\end{figure*}

A major concern for the study of the neutral hydrogen gas--SFR relation is the definition of the galaxy morphology; how we choose what constitutes a disc can affect our results. There are various morphology metrics used for both observations and simulations that allow one to distinguish between discs and ellipticals. However, exploring which one is better is not the task of this work. Here we briefly assess the $\HI$ gas--SFR relation for ellipticals in TNG100 and EAGLE. As shown in Figure~\ref{fig_mhi_sfr_el}, the $\HI$ mass for ellipticals is slightly lower but not significantly different from that of discs in TNG100. This is mainly caused by the higher-than-observed $\HI$ content of TNG100 ellipticals, as pointed out by \cite{Diemer2019}. The independence of the cold gas content at given SFR with morphology is also shown in EAGLE, as seen in Figure~\ref{fig_mhi_sfr_centrals_eagle}. 
These results indicate that the neutral hydrogen gas mass--SFR relation shown in the main text is not significantly affected even if the morphological selection of galaxies in the simulations is not perfect. 

\section{SFR over different time-scales and apertures}
\label{sec_SFR}

\begin{figure*}
\centering
 \includegraphics[width=0.9\linewidth]{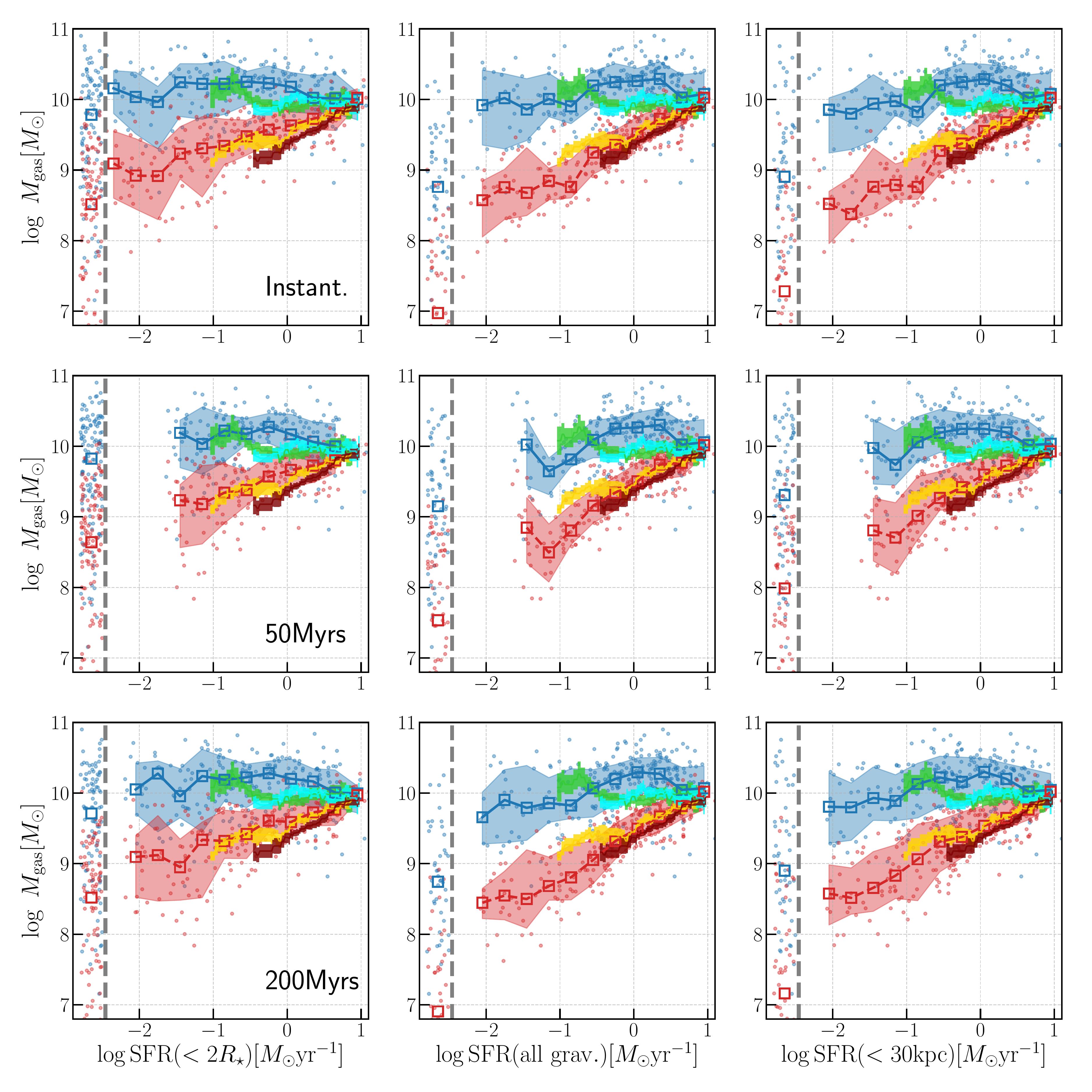}
\caption{\label{fig_timescale_aperture} $\HI$ and ${\rm H_2}$ gas mass versus SFRs calculated over different time-scales and different apertures. TNG100 $\HI$ and ${\rm H_2}$ masses here use the L08 decomposition model. The limegreen and gold lines with error bars are the observational results given in \citetalias{Zhang2019} for $\HI$ and $\H2$ respectively.}
\end{figure*}

In observations, different SFR indicators, such as UV continuum, ${\rm H\alpha}$ emission/$D_{4000}$, IR continuum and $1.4$GHz emission, are sensitive to the SFR over different timescales, varying from less than $\sim 10$\,Myr for ${\rm H\alpha}$ to $\sim 100$\,Myr for UV+IR \citep{2014ApJS..214...15S}. In simulations, SFRs can also be calculated using different time-scales and apertures. To test the impact of time-scale and aperture choices, in Figure \ref{fig_timescale_aperture},  we plot the gas--SFR relation of the L08 model as in Figure \ref{fig_mhi_sfr_disc}, but using SFRs derived within different timescales (from instantaneous to $200$\,Myr) and apertures ($<2R_\star$, $<30$\,kpc, all grav.) \citep{Donnari2019}. There are differences and variations of the results in different panels, but the general trends for both $\HI$ and ${\H2}$ remain similar, and are in general agreement with observations.  We also tested the results derived with other $\HI/\H2$ decomposition models, and find that the general trends remain unchanged against SFRs calculated using different apertures and time-scales. We want to point out that while the time-scale and aperture choices for the simulation are not particularly important, the SFR indicators used for the observations is still very important, as the modeling uncertainties in observations are more complicated than just time-scales and apertures (for example, see the difference between $H_{\alpha}$/$D_{4000}$ and UV+IR SFRs as highlighted in \citetalias{Zhang2019}).

\section{Gas density in Example SF and quenching galaxies}
\label{sec_gas_dens}

\begin{figure*}
\centering
 \includegraphics[width=0.49\linewidth]{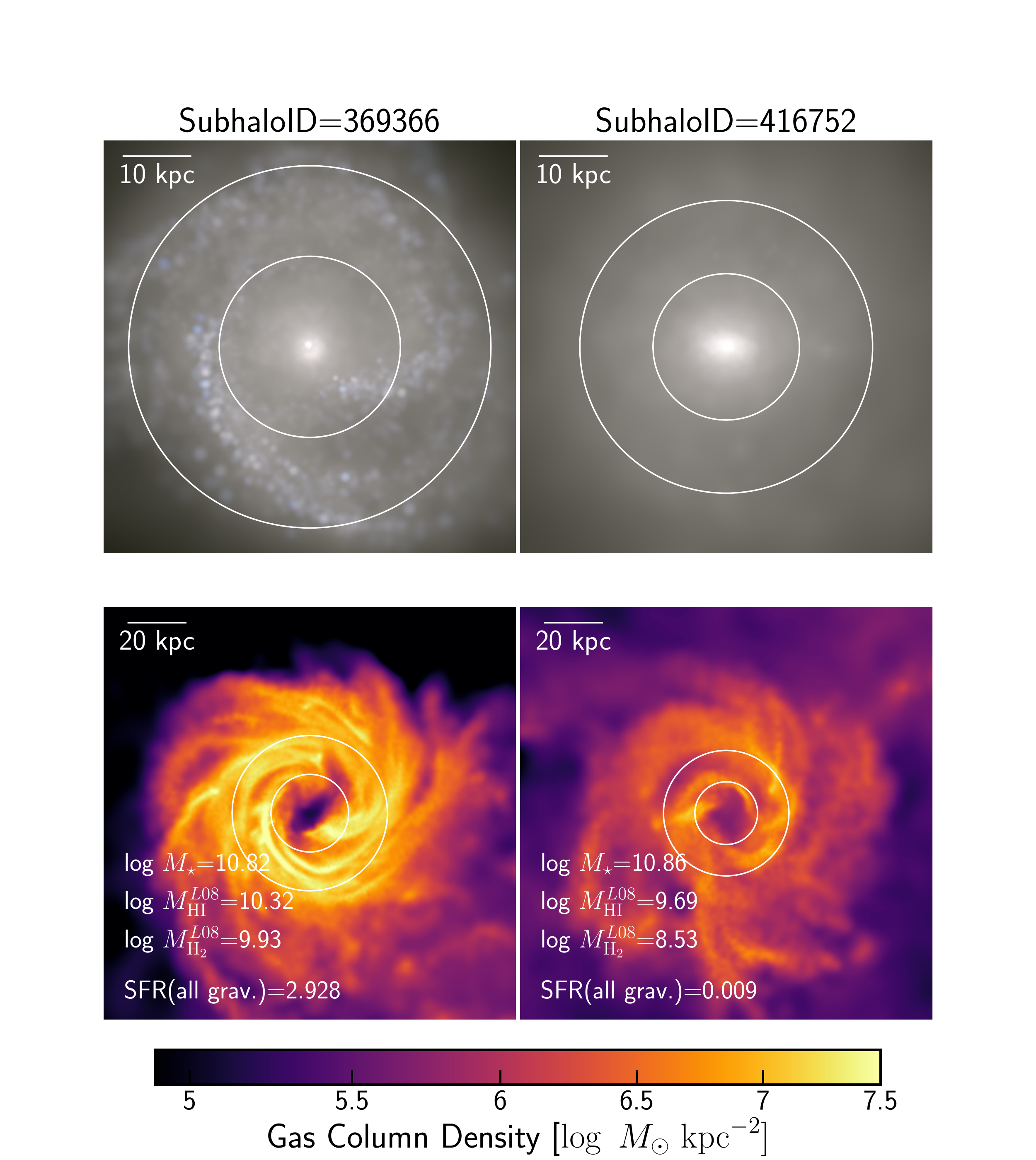}
 \includegraphics[width=.49\linewidth]{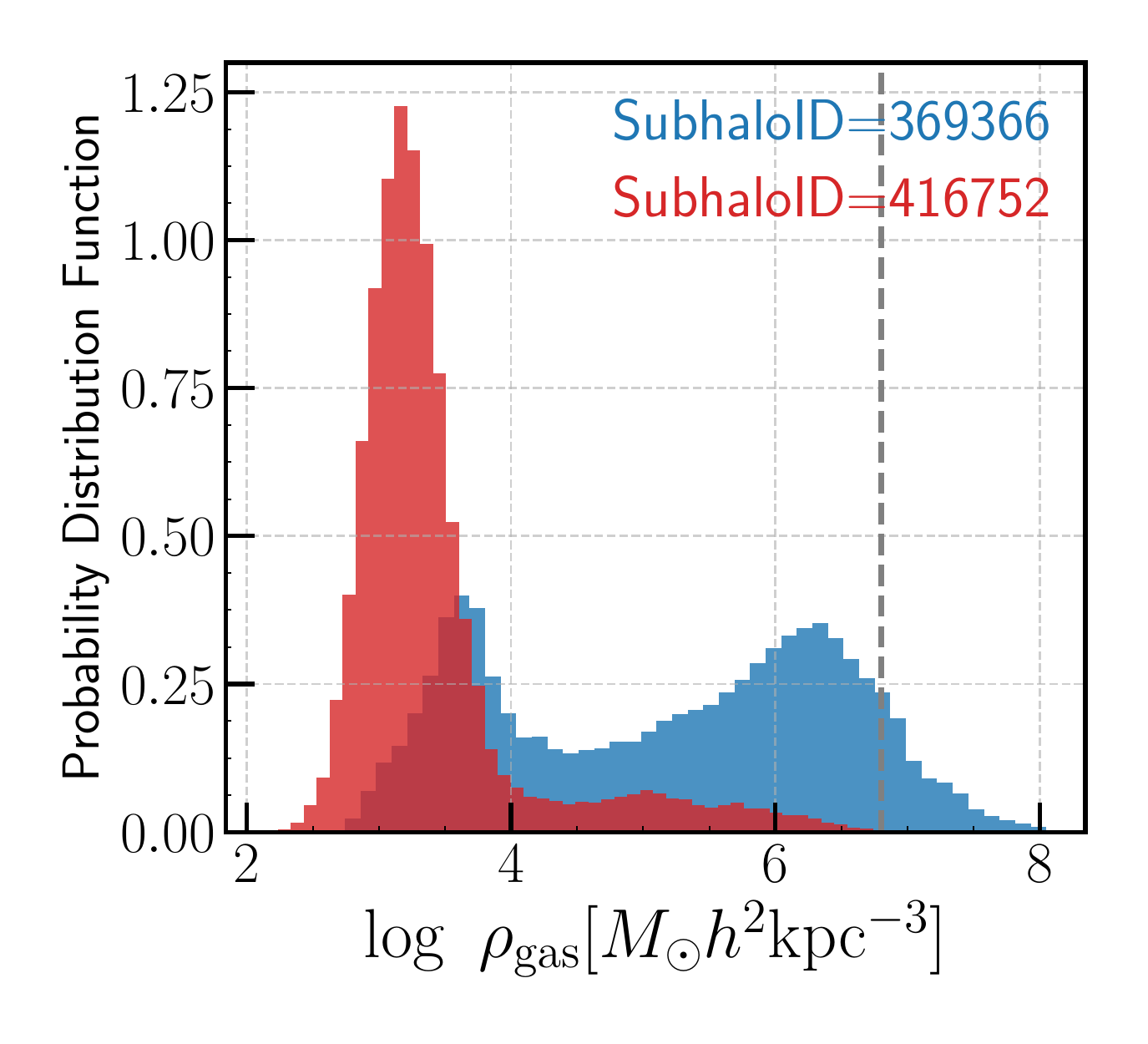}
\caption{\label{fig_gas_dens} 
Left: The composite stellar light of JWST wide bands of F200W+F115W+F070W (upper) and gas (lower) distribution of a representative higher SFR(all grav.)$>10^{-0.5}\msun {\rm yr}^{-1}$ and a lower SFR$_{\rm limit}<$SFR(all grav.)$<10^{-0.5}\msun {\rm yr}^{-1}$ disc galaxy at $z=0$. The images size is $60$\,kpc $\times$ $60$\,kpc for stellar light and $140$\,kpc $\times$ $140$\,kpc for the gas. Galaxies are rotated to be face-on. The circles indicate the $1.0R_{\star}$ and $2.0R_{\star}$ radius separately. Please note that the scale of the color coding for gas density is the same for both galaxies. Masses and SFRs are in units of ${\rm M}_\odot$ and ${\rm M}_\odot\, {\rm yr}^{-1}$, respectively.
Right: The 3D gas density distribution in the example galaxies in TNG100. The gray vertical line indicates the density threshold for star formation adopted in IllustrisTNG model.}
\end{figure*}

In Figure \ref{fig_gas_dens}, we show the stellar light and total gas distribution for a typical  higher SFR central disc galaxy and a lower SFR central disc in TNG100. 
The lower SFR galaxy also shows a regular rotating gas disc, similar to the higher SFR one. This is also in good agreement with the results in \citetalias{Zhang2019}, where they find that most of the quenching central disc galaxies show characteristically symmetric double-horn $\HI$ profiles, indicating regularly rotating $\HI$ discs with little significant kinematic perturbations, similar to the star-forming ones. The difference between the two galaxies is more directly shown in the right panel. 
Comparing to the one with higher SFR, in the lower SFR galaxy, there are fewer regions with high gas density (hence less gas is in the form of ${\H2}$). The dashed line corresponds to density threshold of $0.13\,{\rm cm^{-3}}$ (i.e.~$n_{\rm H}\sim 0.1\,{\rm cm^{-3}}$) set up for star formation. Note the gas cell volumes in quenching galaxy is much larger than the SF one, as a result, the surface density profile of these two galaxies are not significantly different.

\end{document}